\documentclass[aip,pof,amsmath,amssymb,unsortedaddress]{revtex4-1}

\usepackage{graphicx}% Include figure files
\usepackage{dcolumn}% Align table columns on decimal point
\usepackage{color}
\usepackage{bm}% bold math
\usepackage[colorlinks=true,       % false: boxed links; true: colored links
    linkcolor=blue,          % color of internal links (change box color with linkbordercolor)
    citecolor=blue,
    urlcolor=blue]{hyperref}
\usepackage{multirow}
\begin{document}

\title[Precessing sphere]{Shear-driven parametric instability in a precessing sphere}
\author{Yufeng Lin}
\email{yufeng.lin@erdw.ethz.ch}
\affiliation{Institute of Geophysics, ETH Zurich, Sonneggstrasse 5, 8092 Zurich, Switzerland }
\author{Philippe Marti}
\affiliation{Department of Applied Mathematics, University of Colorado Boulder, USA}

\author{Jerome Noir}
\affiliation{Institute of Geophysics, ETH Zurich, Sonneggstrasse 5, 8092 Zurich, Switzerland }

\date{\today}% It is always \today, today,

\begin{abstract}
The present numerical study aims at shedding light on the mechanism underlying the precessional instability in a sphere. Precessional instabilities in the form of parametric resonance due to topographic coupling have been reported in a spheroidal geometry both analytically and numerically.  We show that such parametric resonances can also develop in spherical geometry due to the conical shear layers driven by the Ekman pumping singularities at the critical latitudes. Scaling considerations lead to a stability criterion of the form, $|P_o|>O(E^{4/5})$, where $P_o$ represents the Poincar\'e number and $E$ the Ekman number. The predicted threshold is consistent with our numerical simulations as well as previous experimental results. When the precessional forcing is supercriticial, our simulations show evidence of an inverse cascade, i.e. small scale flows merging into large scale cyclones with a retrograde drift. Finally, it is shown that this instability mechanism may be relevant to precessing celestial bodies such as the Earth and Earth's moon. 
\end{abstract}

\keywords{precession, shear layer, parametric instability}

\maketitle

\section{Introduction}\label{Sec:Intro}
Precession corresponds to the gyroscopic motion of a rotating object due to a torque orthogonal to its spin axis.  When an internal liquid layer is present, such as the liquid core or the subsurface ocean of a planet or the fuel tank of a spinning spacecraft, energy can be dissipated in the liquid due to the induced flows. In line with this idea, it has been proposed that orbital perturbations, in particular precession and nutation, could be used to probe the interior of the planet \cite{Thomson1882,Poincare1910,Williams2001} and possibly generate magnetic fields through the so-called dynamo process \cite{Bullard1949,Tilgner2005,Tilgner2007a,Wu2009,Dwyer2011}.   

It is well-established that the primary response in the bulk of a rapidly rotating fluid cavity subject to precession is of uniform vorticity, i.e. the fluid rotates along an axis tilted with respect to the mean rotation axis of the container \cite{Poincare1910,Busse1968,Noir2003,Noir2013,Zhang2014}. Although viscosity plays an important role mostly in the viscous boundary layer next to the solid wall of the cavity, it has been shown that singularities at the so-called critical latitudes excite conical shear layers in the interior that are superimposed on the uniform vorticity flow \cite{Bondi1953,Stewartson1963,Kerswell1995,Kida2011}. In addition, weakly non-linear interactions in these singular regions of the boundary layer generate steady geostrophic shears \cite{Busse1968}, which were observed both numerically \cite{Hollerbach1995, Tilgner2001, Noir2001} and experimentally \cite{Malkus1968, Vanyo1995, Noir2001b, Noir2003, Goto2007,Triana2012,Boisson2012}. 

In a pioneering piece of experimental work, Malkus \cite{Malkus1968} showed that precession driven flows can become unstable and even create space filling turbulence. These initial findings were re-confirmed by later experiments \cite{Vanyo1995,Noir2003,Goto2007,Goto2014}. A theoretical justification was first proposed by Kerswell \cite{Kerswell1993}, who argued that the precession driven uniform vorticity flow in a spheroidal cavity is inertially unstable due to the constant strain field exerted by the solid wall. Indeed, in a spheroid, the uniform vorticity flow is a superposition of a solid body rotation and a gradient flow, which is required to fulfil the non-penetration condition at the wall. This gradient flow can be decomposed into two components, one leading to elliptically deformed streamlines and the other to a shear of the centres of the streamlines. Both parts can interact with a pair of free inertial modes of the rotating cavity through parametric couplings. This mechanism has been confirmed numerically \cite{Lorenzani2003} for oblate spheroids,  while similar inertial instabilities were reported experimentally in cylindrical precessing tanks \cite{Lagrange2008,Lagrange2011,Lin2014}. 

Both numerical simulations \cite{Tilgner2001, Hollerbach2013,Wei2013} and laboratory experiments \citep{Goto2007,Goto2014} in spherical geometries have shown very rich dynamics ranging from laminar flows to fully developed turbulence. However, in the case of a spherical cavity the uniform vorticity solution reduces to a purely solid body rotation, preventing the aforementioned inertial instability to develop. Hence, instabilities in a sphere can only originate from the viscous correction to the solid body rotation flow \citep{Lorenzani2001}.  

The present study aims at shedding light on the onset of the unstable flows in a precessing sphere. Our numerical results show that the shear in the conical structures spawned from the critical latitudes is key to the destabilization process, echoing the shear of the centres of the streamlines in the case of a spheroidal cavity. We conjecture that the conical shear layers couple with two free inertial modes leading to a parametric resonance. The known scalings of the conical shear layers allow us to derive a stability criterion for the onset of the unstable flow that is in agreement with our numerical simulations. Furthermore, we show that for large enough forcing small scale vortices merge into large scale cyclonic vortices with a retrograde drift. 
 
The paper is organized as follows. We first introduce the mathematical background and the numerical model in Sec. \ref{Sec:Math}, and then we present our numerical results in Sec. \ref{Sec:Resu}. Using heuristic arguments, we derive a scaling for the onset of the observed instability in Sec. \ref{Sec:Scal}. Finally, we discuss our findings in the context of planetary dynamics in Sec. \ref{Sec:Diss}.   

\section{Mathematical background and numerical method}\label{Sec:Math}
\begin{figure}
\begin{center}
\includegraphics[width=0.5\textwidth]{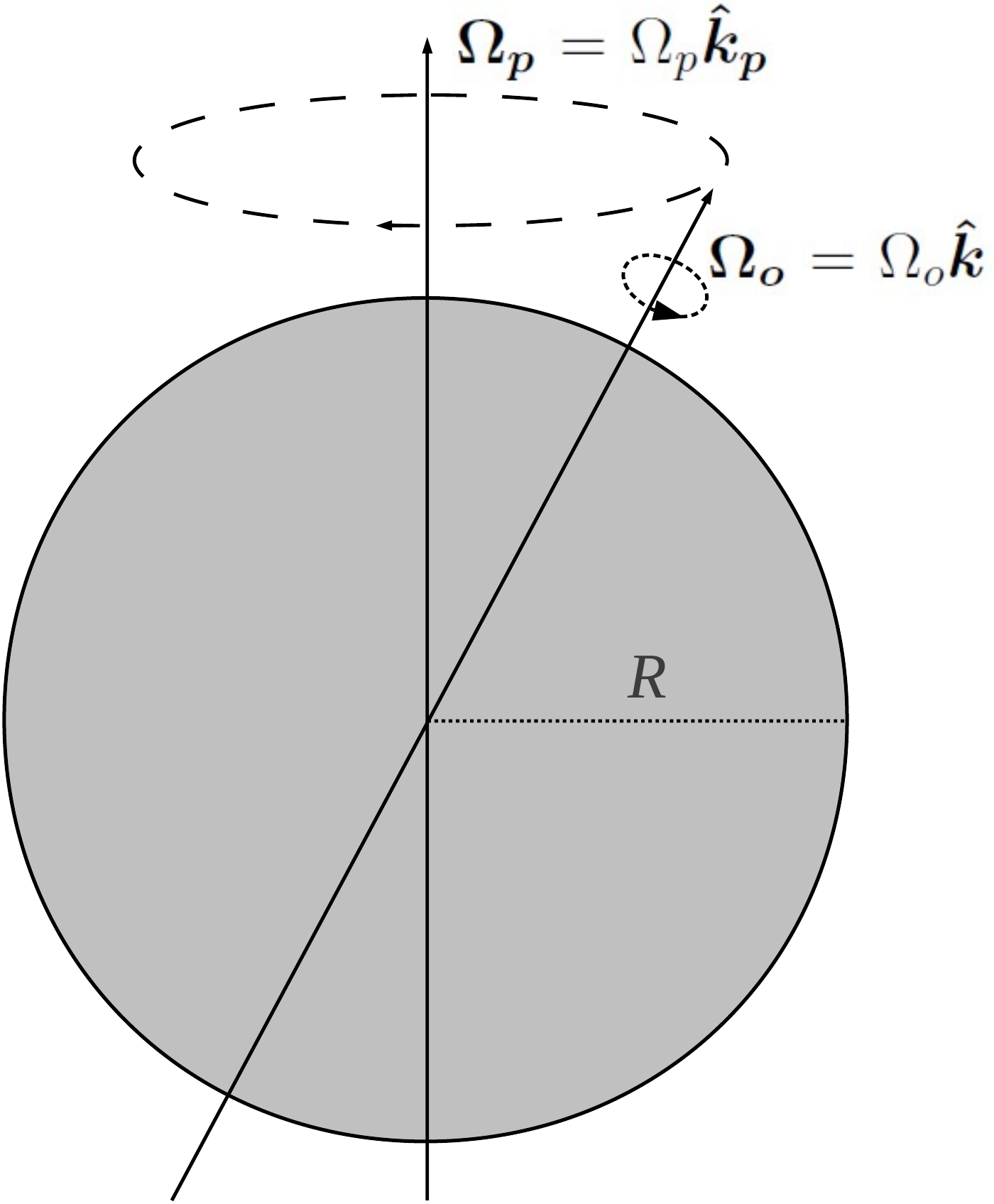}
\end{center}
\caption{Sketch of the problem.}
\label{fig:problem}
\end{figure}

We consider a sphere of radius $R$ filled with a homogeneous and incompressible fluid of density $\rho$ and kinematic viscosity $\nu$. The sphere rotates at $\bm{\Omega_o}=\Omega_o\bm{\hat{k}}$ and precesses at $\bm{\Omega_p}=\Omega_p\bm{\hat{k}_p}$, where $\bm{\hat{k}}$ and $\bm{\hat{k}_p}$ are unit vectors along the spin and precession axes, respectively (Figure \ref{fig:problem}). Hereinafter we refer to the mantle frame as the frame attached to the container, the precession frame as the frame rotating at $\bm{\Omega_p}$ and the fluid frame, the frame attached to the solid body rotation of the fluid. In this paper, all the numerical calculations are implemented in the mantle frame, while the precession frame and the fluid frame are used from time to time to discuss the results.  
        
\subsection{Equations}
Using the radius $R$ as the length scale and $\Omega_o^{-1}$ as the time scale, the dimensionless Navier-Stokes equations describing the fluid velocity $\bm u$ and pressure $p$ in the mantle frame become \cite{Zhang2010a}
\begin{equation}
\frac{\partial \bm u}{\partial t}+\bm u\cdot \nabla \bm u+2(\bm{\hat{k}}+P_o \bm{\hat{k}_p})\times \bm u=-\nabla p+E\nabla^2\bm u-P_o(\bm{\hat{k}_p}\times \bm{\hat{k}})\times \bm r,
\label{eq:NS1}
\end{equation}
\begin{equation}
\nabla\cdot \bm{u}=0,
\label{eq:NS2}
\end{equation}
where the time-dependent precession vector $\bm{\hat{k}_p}$ is given by
\begin{equation}
\bm{\hat k_p}=(\bm{\hat \imath}\cos t-\bm{\hat \jmath}  \sin t) \sin \alpha_p+\bm{\hat k} \cos \alpha_p.
\label{eq:kp}
\end{equation}
Here $(\bm{\hat \imath},\bm{\hat \jmath},\bm{\hat k})$ are unit vectors of  Cartesian coordinates $(x,y,z)$ whose $z$-axis is along the rotation vector $\bm{\hat k}$. In Eq. (\ref{eq:NS1}), the last term on the left hand side is the Coriolis force due to the rotation and precession, the last term on the right hand side is the so-called Poincar\'e acceleration and $p$ is the reduced pressure.

In the present study the angle $\alpha_p$ between the spin and precession axes is kept constant and equal to $60^{\circ}$. The evolution of the system is then governed by two dimensionless parameters, the Poincar\'e number $P_o$ and the Ekman number $E$
\begin{equation}
P_o=\frac{\Omega_p}{\Omega_o}, \quad E=\frac{\nu}{\Omega_o R^2},
\end{equation}
which measure the dimensionless rate of precession and the ratio of the typical viscous force and Coriolis force, respectively. Negative (positive) values of $P_o$ correspond to retrograde (prograde) precession; we consider only retrograde precession in the present study. 

\subsection{Numerical solver}
Equations (\ref{eq:NS1}) and (\ref{eq:NS2}), together with no slip boundary condition $\bm u=0$ on the wall, are solved using the fully spectral code developed by P. Marti \cite{MartiThesis2012}. Using the so-called toroidal/poloidal decomposition in a spherical coordinate system $(r,\theta,\phi)$ 
\begin{equation}
\bm u=\nabla\times(T \bm{r})+\nabla\times\nabla\times(P \bm{r}),
\label{eq:TPD}
\end{equation}
 the velocity field is then represented by two scalar fields $T$ and $P$. The incompressibility condition is automatically satisfied by such a decomposition. The scalar fields $T$ and $P$ are then expanded as
 \begin{equation}
 T(r,\theta,\phi)=\sum_{n=0}^{N}\sum_{l=0}^L\sum_{m=-l}^{l} t_{l,n}^mW_n^l(r)Y_l^m(\theta,\phi),
\label{eq:TEx}
\end{equation}  
 \begin{equation}
 P(r,\theta,\phi)=\sum_{n=0}^{N}\sum_{l=0}^L\sum_{m=-l}^{l} p_{l,n}^mW_n^l(r)Y_l^m(\theta,\phi),
\label{eq:PEx}
\end{equation} 
where $Y_l^m(\theta,\phi)$ are the spherical harmonics of degree $l$ and order $m$ and $W_n^l(r)$ are the so-called Worland polynomials. For a given harmonic degree $l$, the Worland polynomials are  a combination of the $r^l$ factor and the Jacobi polynomials $P_n^{(\alpha,\beta)}(x)$, i.e.  $W_n^l(r)=r^l P_n^{-1/2,l-1/2}(2r^2-1)$, which exactly satisfy the parity and regularity at the origin of the sphere \cite{Livermore2007}. 

Substituting Eq. (\ref{eq:TPD}-\ref{eq:PEx}) into Eq. (\ref{eq:NS1}) and taking the $r$-component of the curl and the curl of the curl yield a set of time evolution equations of the spectral coefficients $t_{l,n}^m$ and $p_{l,n}^m$. The time integration is implemented using a second order predictor-corrector scheme. 

The numerical code has been widely benchmarked in several contexts including that of precession driven flows \citep{Hollerbach2013,Marti2014}. 
 We use a typical truncation of the spectral expansion up to $N=63$, $L=M=127$ at moderate Ekman numbers ($E\geqslant3\times 10^{-5}$). For a few calculations at lower Ekman numbers ($E\leqslant 10^{-5}$), we truncated as high as $N=255$, $L=511$ and $M=255$ to well resolve the thin boundary layer and instabilities.  

\subsection{Derived quantities}
In this subsection, we introduce some useful quantities derived from our numerical simulations performed in the mantle frame.

The solution to the linearized equations (\ref{eq:NS1}) and (\ref{eq:NS2}) is symmetric around the origin due to the parity of the precessional forcing \citep{Zhang2010a,Hollerbach2013} (see also in Appendix \ref{app_symmetry}), namely $\bm{u}(-\bm r)=-\bm u(\bm r)$, where $\bm r$ is the position vector. Any breaking of the symmetry must be caused by an instability. Hence, it is of interest to decompose the total velocity into its symmetric and antisymmetric parts, i.e. $\bm u=\bm u_s+\bm u_a$ with
\begin{equation}
\bm u_s=\frac{\bm{u}(\bm r)-\bm u(-\bm r)}{2}, \quad \bm u_a=\frac{\bm{u}(\bm r)+\bm u(-\bm r)}{2}.
\end{equation}
Such a decomposition can easily be  performed in the spectral space by taking advantage of the parity of the spherical harmonics. Specifically, the symmetric part includes only odd degree $l$ in the toroidal field $T$ and even degree $l$ in the poloidal field $P$, while the antisymmetric part is the other way round. 

Following the symmetry decomposition, we define the total energy, symmetric energy and antisymmetric energy as:
\begin{equation}
Ek=\frac{1}{2}\int_V |\bm u(\bm r)|^2 \mathrm d V,\quad
Ek_s=\frac{1}{2}\int_V |\bm u_s(\bm r)|^2 \mathrm d V, \quad
Ek_a=\frac{1}{2}\int_V |\bm u_a(\bm r)|^2 \mathrm d V.
\end{equation}

The growth of an anti-symmetric energy component is a sufficient condition to identify the development of an instability. However, it should be noted that a symmetry-breaking is not a necessary condition, indeed instabilities with negligible anti-symmetric component have been reported in numerical studies \cite{Lorenzani2003}. In all cases, the instabilities will lead to time variations of the energy in the system, that is otherwise steady.

The main observations from previous experimental studies are reported in the precession frame \cite{Malkus1968,Vanyo1995, Noir2003}, it is therefore natural to provide the reader with an equivalent point of view in our numerical simulations. 
To that end, when needed, after we perform our simulations in the mantle frame, velocities are simply transformed in the precession frame as
\begin{equation}
\bm{u}^p=\bm{u}+\hat{\bm{z}}\times\bm{r}.
\end{equation}

Anticipating the peculiar role of the conical shear layers which are coaxial with the rotation axis of the fluid\cite{Vanyo1995,Noir2003}, we extract the rotation vector of the fluid $\bm{\omega_F}$  from the mean vorticity in the bulk \cite{Tilgner2001}
\begin{equation}
2\bm {\omega_F}=<\nabla\times\bm{u}^P>,
\end{equation} 
where  $<\cdot>$ denotes averaging in the fluid volume excluding a thin boundary layer ($10E^{1/2}$). 

Once $\bm {\omega_F}$ is determined, the spectral coefficients with respect to the system of coordinates aligned with the rotation axis of the container are converted into a new set associated with the coordinate system aligned with $\bm {\omega_F}$, from which the velocities are reconstructed. This transformation is made using a Matlab subroutine \cite{Simons2014} 
%developed by F. Simon \footnote{\url{http://geoweb.princeton.edu/people/simons/software.html}} 
based on the formula in Ref. \onlinecite{Dahlen1998}. The angle $\alpha_f$ between the rotation axes of the container and the fluid is defined as $\cos{\alpha_f}=\bm{\hat{k}}\cdot \bm{\omega_F}$. Finally, $\varepsilon=|\bm{\hat{k}}-\bm{\omega_F}|$ represents the differential rotation between the container and the bulk of the fluid. 
%Note that $\varepsilon\approx\alpha_f$ provided $\alpha_f \ll 1$.

\section{Results}\label{Sec:Resu}
\subsection{Base flow}
\begin{figure}
\begin{center}
\includegraphics[width=0.48 \textwidth]{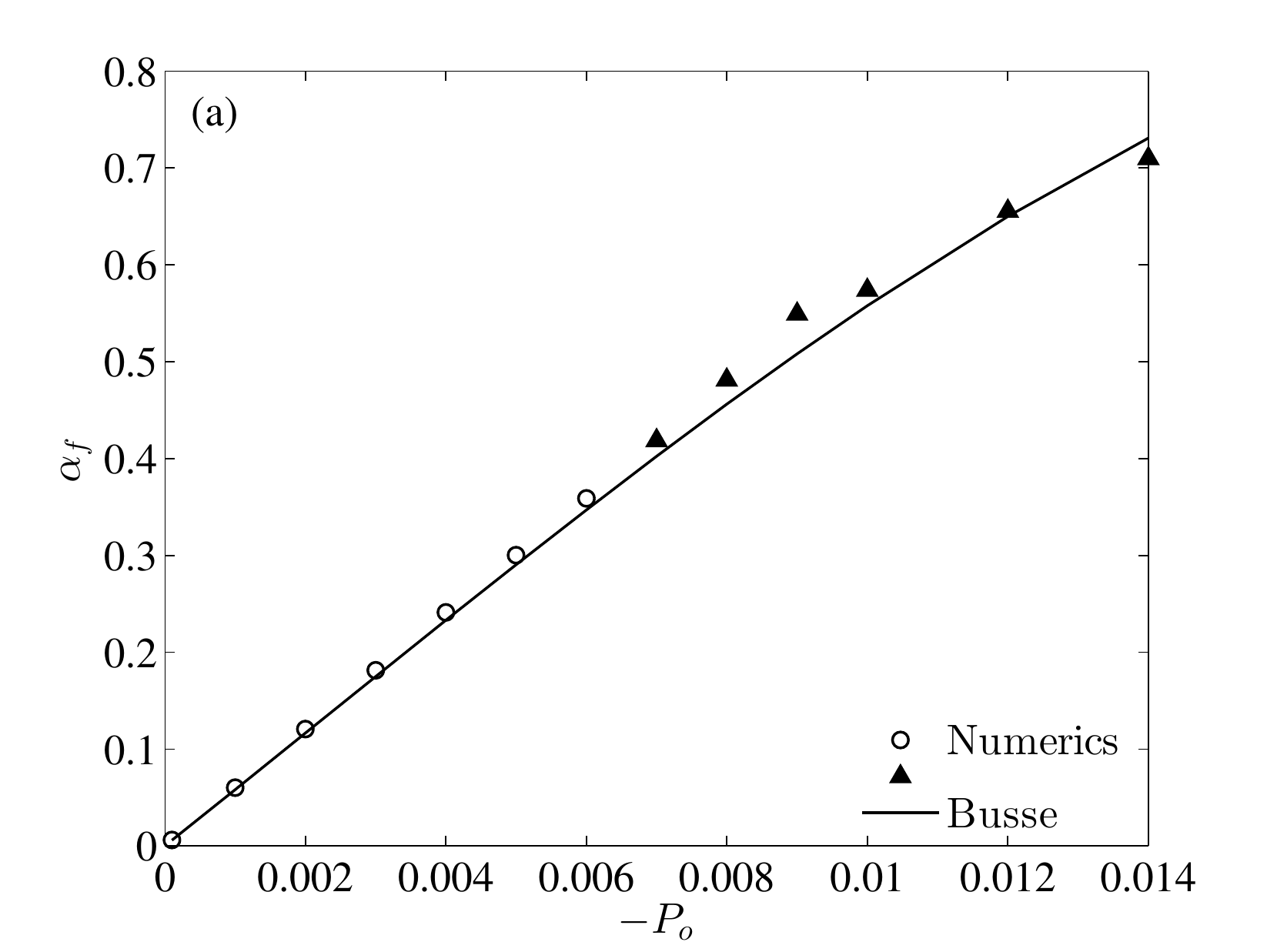}
\includegraphics[width=0.48 \textwidth]{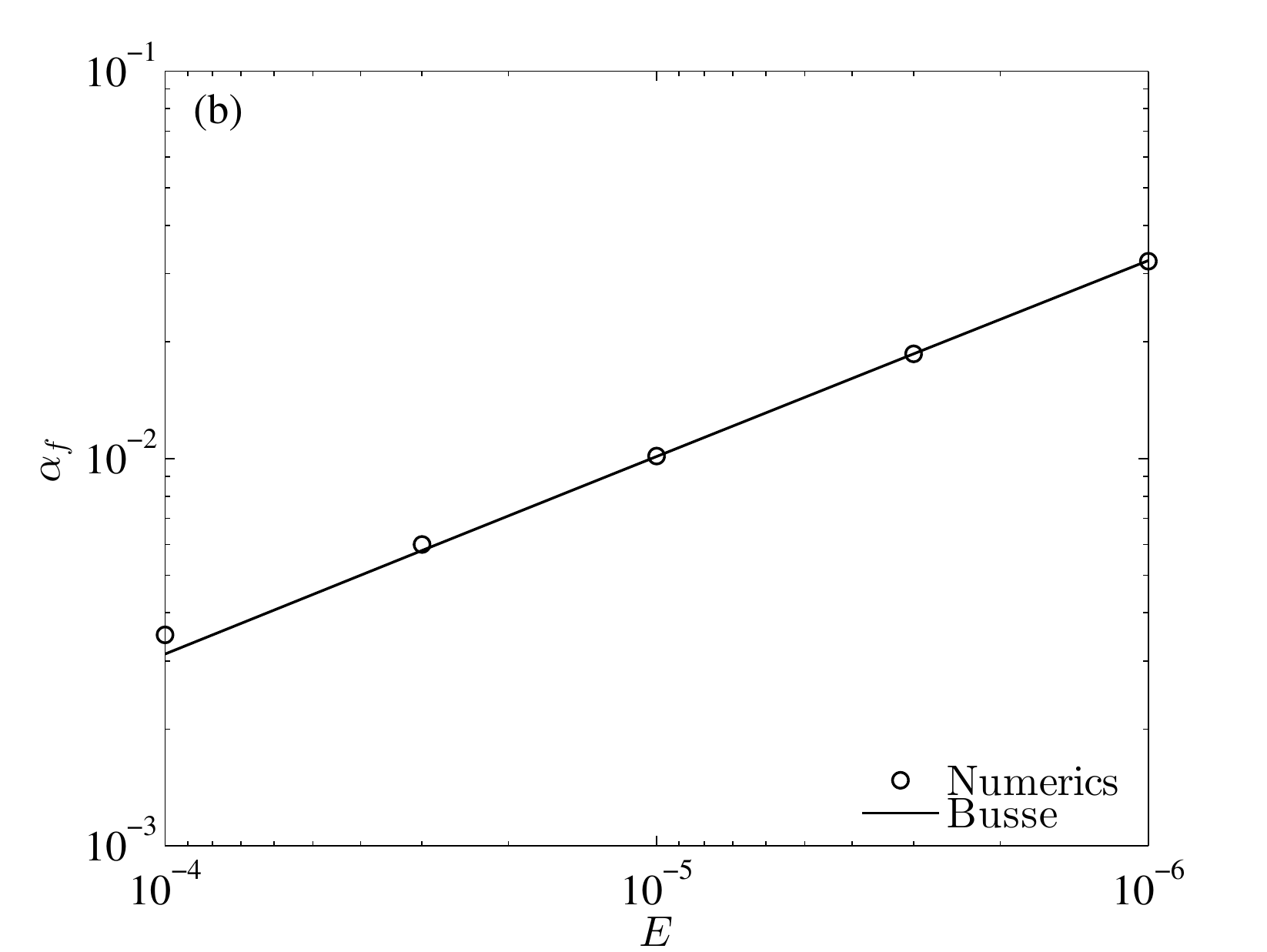}
\end{center}
\caption{The angle $\alpha_f$ (in radians) between rotation axes of the container and the fluid    (a) as a function of the Poincar\'e number $P_o$ at fixed Ekman number $E=3.0\times 10^{-5}$, and (b) as a function of the Ekman number $E$ at fixed $Po=-1.0\times 10^{-4}$. The solid lines represent Busse's theory \cite{Busse1968} and symbols correspond to our numerical results. The open circles ($\circ$) and the solid triangles ($\blacktriangle$) represents  stable and unstable flows, respectively.}
\label{fig:alpha_f}
\end{figure}

\begin{figure}
\begin{center}
\includegraphics[width=0.9 \textwidth]{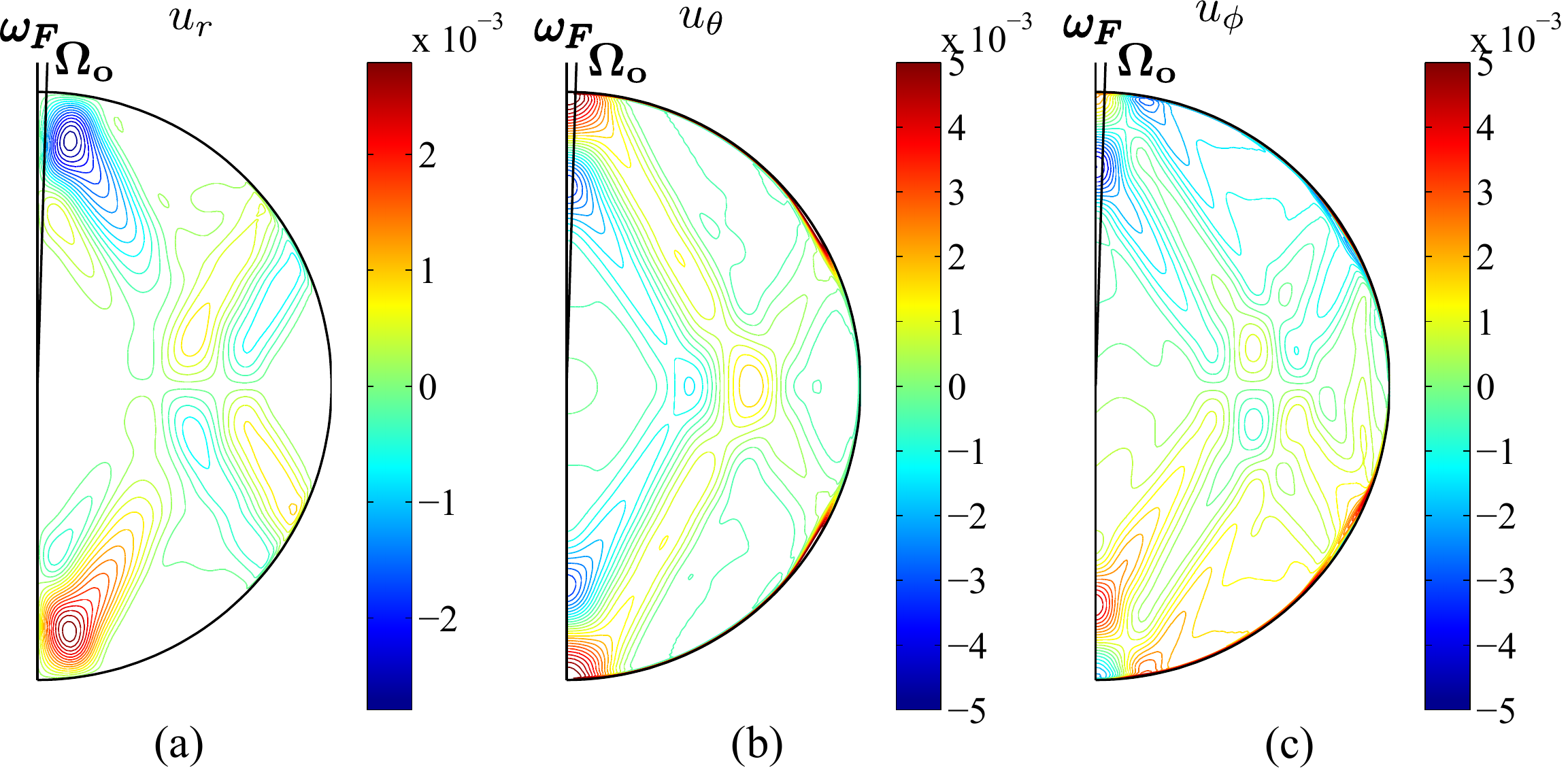}
\end{center}
\caption{Contours of the velocities in the fluid frame in the meridional plane across both $\bm{\omega_F}$ and $\bm {\Omega_o}$ at $P_o=-1.0\times 10^{-4}$ and $E=1.0\times 10^{-6}$.}
\label{fig:conical}
\end{figure}

It is now well established that the base flow in a weakly precessing sphere is in the form of a solid body rotation along an axis inclined to the rotation axis of the container \cite{Busse1968,Zhang2010a,Kida2011}. Figure \ref{fig:alpha_f} represents the angle $\alpha_f$ between the rotation axes of the container and the fluid as a function of the Poincar\'e number $P_o$ and Ekman number $E$. We observe a quantitative agreement between our simulations (symbols) and the theoretical predictions\cite{Busse1968} (solid lines) that have been thoroughly validated \citep{Tilgner2001,Noir2003}. We note in particular that the scaling $\alpha_f \propto E^{-1/2}$ at fixed $P_o$ expected from the resonance at $|P_o|\cos \alpha_p \ll E^{1/2}$ is well recovered \citep{Busse1968}. Figure \ref{fig:alpha_f} also includes data points for which an instability occurs ($\blacktriangle$), showing that even in these cases the asymptotic theory of Busse \cite{Busse1968} remains valid at first order. 

A dominant feature of the viscous corrections to the solid body rotation flow is the conical shear layers spawned from the critical latitudes. These structures, coaxial with $\bm{\omega_F}$,  correspond to the so-called characteristic surfaces of the hyperbolic inertial wave equation, i.e. the unforced inviscid Navier-Stokes equations in the rotating frame. The evidence of such oblique shear layers has been reported in numerical and experimental studies \citep{Hollerbach1995, Noir2001, Noir2001b}, with a typical velocity of  $O(\varepsilon E^{1/5})$ over a typical width of $O(E^{1/5})$ \cite{Stewartson1963,Kerswell1996,Kida2011}. A typical example at $Po=-1.0\times 10^{-4}$ and $E=1.0\times 10^{-6}$ is represented in Figure \ref{fig:conical}, showing the conical shear layers in the fluid frame. As we shall see later on in this paper, they play an important role in the destabilization mechanism of the flow.  

\subsection{The parametric instability regime}
\begin{figure}
\begin{center}
\includegraphics[width=0.8 \textwidth]{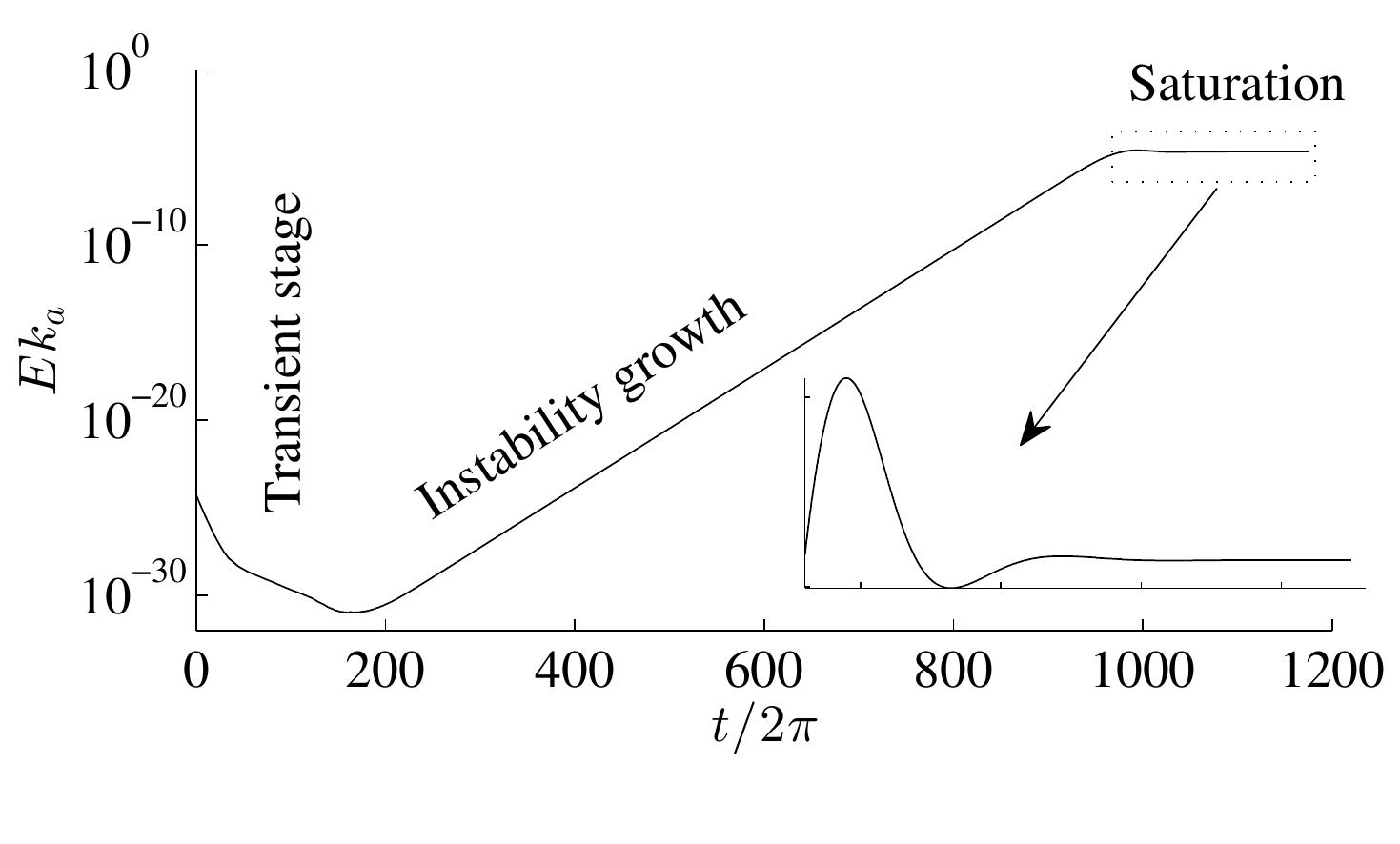}
\caption{Anti-symmetric energy $Ek_a$ as a function of time in the mantle frame. $P_o=-7.0\times 10^{-3}$, $E=3.0\times 10^{-5}$.}
\label{fig:Vel_Ek_E3e-5P7e-3}
\end{center}
\end{figure}

\begin{figure}
\begin{center}
\includegraphics[width=15cm]{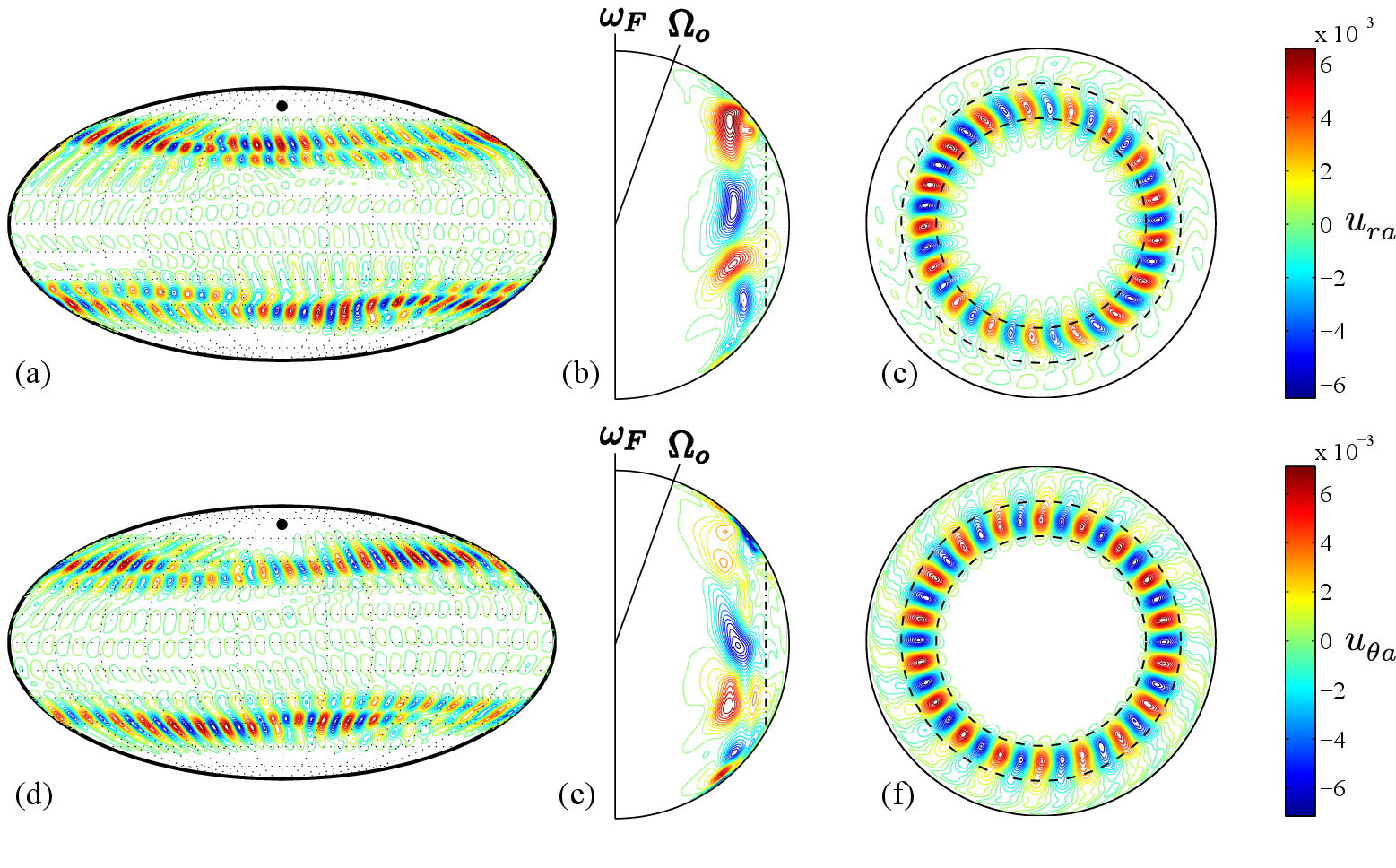}
\caption{Contours of the antisymmetric velocities $u_{ra}$ (a-c) and $u_{\theta a}$ (d-f) in the fluid frame at $t/2\pi=973$, $P_o=-7.0\times 10^{-3}$ and $E=3.0\times 10^{-5}$. The left column is on a spherical surface of $r=1-10E^{1/2}$ and the black dot denote the rotation axis of the container. The middle column is in the meridional plane across both $\bm{\omega_F}$ and $\bm {\Omega_o}$. The vertical dashed line in the meridional plane represents the cylinder associated with the critical latitude. The right column is in the equatorial plane perpendicular to the rotation axis of fluid. Two dashed circles are $r=0.6$ and $r=0.8$.}
\label{fig:ua_E3e-5P7e-3}
\end{center}
\end{figure}    

\begin{figure}
\begin{center}
\includegraphics[width=10cm]{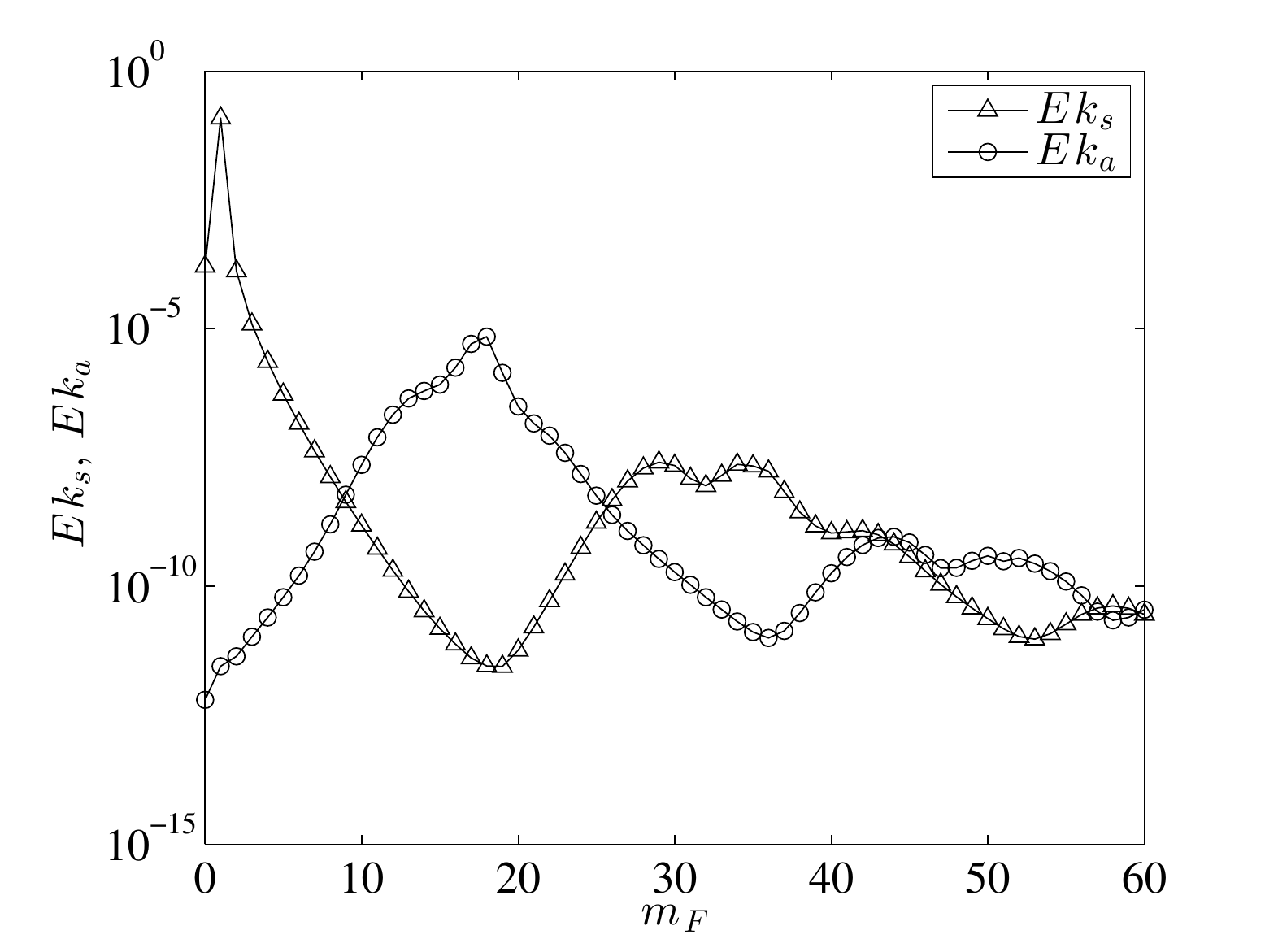}
\caption{Symmetric energy $Ek_s$ ($\triangle$) and antisymmetric energy $Ek_a$ ($\circ$) in individual $m_F$ modes for the same instant and parameters as in figure \ref{fig:ua_E3e-5P7e-3}.}
\label{fig:spectrum_mF_E3e-5P7e-3}
\end{center}
\end{figure} 

As we increase the precession rate at a given Ekman number $E$, the base flow becomes unstable. A typical example just above the threshold is presented in Figs. \ref{fig:Vel_Ek_E3e-5P7e-3} - \ref{fig:spectrum_mF_E3e-5P7e-3} for $P_o=-7.0\times 10^{-3}$ and $E=3.0\times 10^{-5}$. Figure \ref{fig:Vel_Ek_E3e-5P7e-3} represents the time evolution of the antisymmetric energy $Ek_a$ in the mantle frame. After a transient stage, $0<t/2\pi<200$, the antisymmetric energy $Ek_a$ grows exponentially until saturation is reached at $t/2\pi \sim1000$.

Figure \ref{fig:ua_E3e-5P7e-3} shows a snapshot ($t/2\pi=973$, just before the saturation) of the antisymmetric velocities $u_{ra}$ and  $u_{\theta a}$ in the fluid frame, on a spherical shell at $10{E^{1/2}}$ below the surface (a,d), in a meridional cross section (b,e) and in the equatorial plane perpendicular to $\bm{\omega_F}$ (c,f). The antisymmetric flow is mostly confined between two cylinders ($0.6<s<0.8$) co-axial with $\bm{\omega_F}$, which correspond to latitudes higher than the critical latitudes in the Ekman layer at $\pm 30^{\circ}$. In the equatorial plane perpendicular to $\bm{\omega_F}$, the antisymmetric velocities are dominated by $m_F=17$ for $u_{ra}$ and $m_F=18$  for $u_{\theta a}$, where $m_F$ is the azimuthal wavenumber with respect to $\bm{\omega_F}$. This is consistent with the symmetry of the velocity components, indeed for an antisymmetric flow: 
\begin{equation}
u_{ra}(\bm r)=-u_{ra}(-\bm r), \quad u_{\theta a}(\bm r)=u_{\theta a}(-\bm r), \quad \quad u_{\phi a}(\bm r)=-u_{\phi a}(-\bm r). 
\end{equation} 
Hence, only modes with odd $m_F$ contribute to $u_{ra}$ and modes with even $m_F$ to $u_{\theta a}$ in the equatorial plane.

Figure \ref{fig:spectrum_mF_E3e-5P7e-3} shows the contribution of individual $m_F$ modes to the symmetric and antisymmetric energy at the same instant and parameters as in figure \ref{fig:ua_E3e-5P7e-3}. The symmetric energy $Ek_s$ is dominated by the $m_F=1$ component, whereas the antisymmetric energy is dominated by the $m_F=17$ and $m_F=18$ components. These observations in the spectrum are consistent with the observed velocities in figure \ref{fig:ua_E3e-5P7e-3}. The less significant peaks at higher wavenumbers may be attributed to secondary bifurcations \citep{Lorenzani2001}.

\begin{figure}
\begin{center}
\includegraphics[width=10cm]{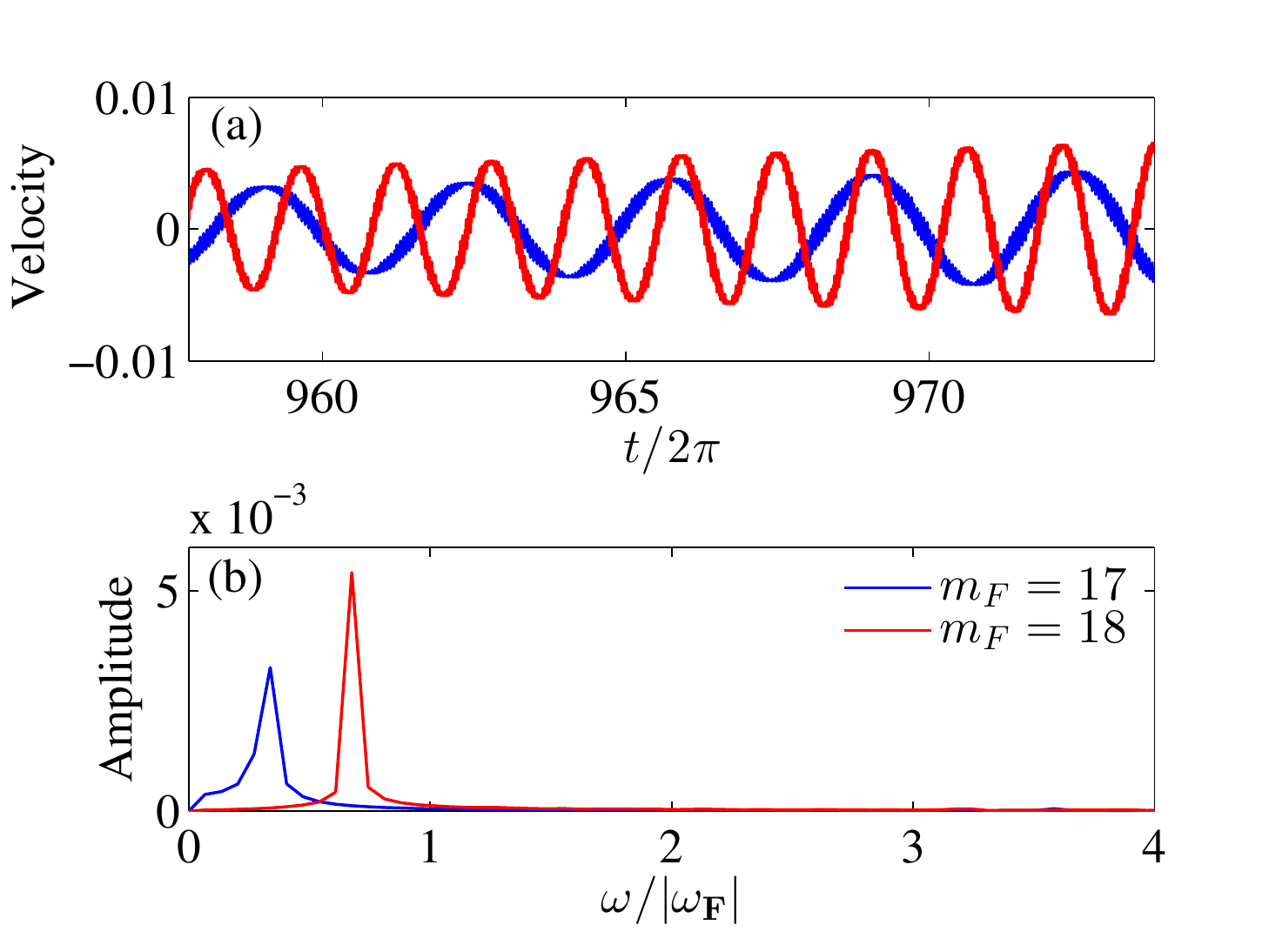}
\end{center}
\caption{Time series of $u_{r}$ of $m_F=17$ mode (in blue) and $u_{\theta}$ of $m_F=18$ mode (in red) at a fixed position ($r=0.7, \theta=\pi/2, \phi=0$) in the fluid frame, and (b) the corresponding discrete Fourier transform. $P_o=-7.0\times 10^{-3}$, $E=3.0\times 10^{-5}$.}
\label{fig:vel_a_FFT_E3e-5P7e-3}
\end{figure}

In figure \ref{fig:vel_a_FFT_E3e-5P7e-3} (a), we extract time series of the velocities corresponding to $m_F=17$ and $m_F=18$  at a fixed position in the fluid frame during the growth phase of the antisymmetric energy. The associated Discrete Fourier Transform (DFT) is shown in figure \ref{fig:vel_a_FFT_E3e-5P7e-3} (b) where frequencies are normalized by $|\bm {\omega_F}|$. The animation (Movie 1 in supplemental material \cite{Note1}) of the velocities in the equatorial plane indicates that the $m_F=17$ mode is prograde and the $m_F=18$ mode is retrograde. So the frequencies of the two modes are  ${\omega_{17}}/{|\bm{\omega_F}|}=-0.34$ and ${\omega_{18}}/{|\bm{\omega_F}|}=0.67$ respectively, which satisfy ${\omega_{18}}/{|\bm{\omega_F}|}-{\omega_{17}}/{|\bm{\omega_F}|}\approx 1.0$. We also see some small amplitude high frequency fluctuations in the time series which may be due to uncertainties in the determination of $\bm {\omega_F}$.

\begin{figure}
\begin{center}
\includegraphics[width=10cm]{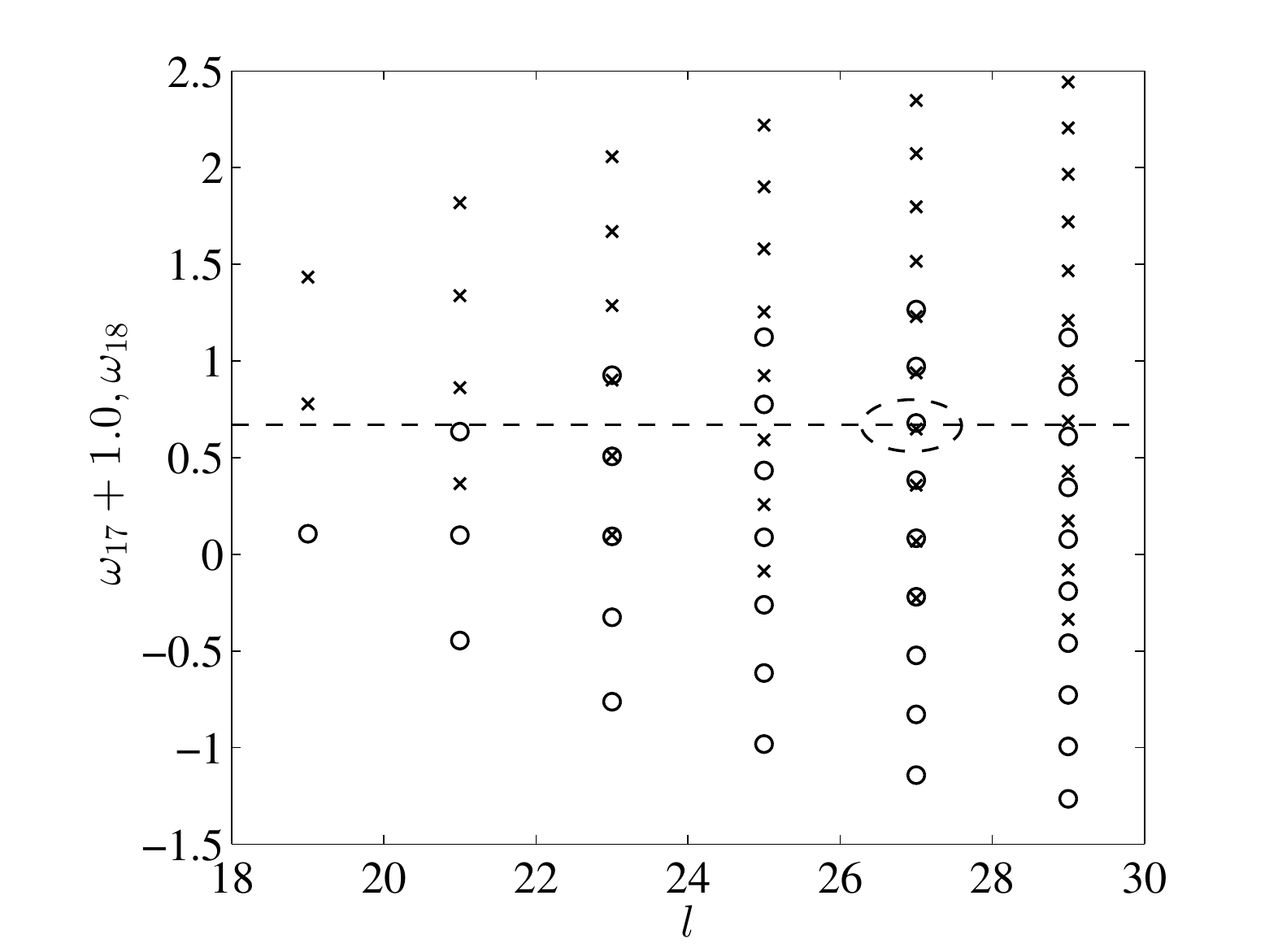}
\end{center}
\caption{Eigen frequencies of inertial modes with $m=17$ ($\times$) and $m=18$ ($\circ$) as a function of modal degree $l$. The horizontal dashed line indicates the observed frequency $\omega_{18}$. A combination in the dashed ellipse indicates a pair of modes closely matching the observations.}
\label{fig:eig_freq}
\end{figure}

\begin{figure}
\begin{center}
\includegraphics[width=7cm,clip,trim=2cm 1cm 2cm 0cm]{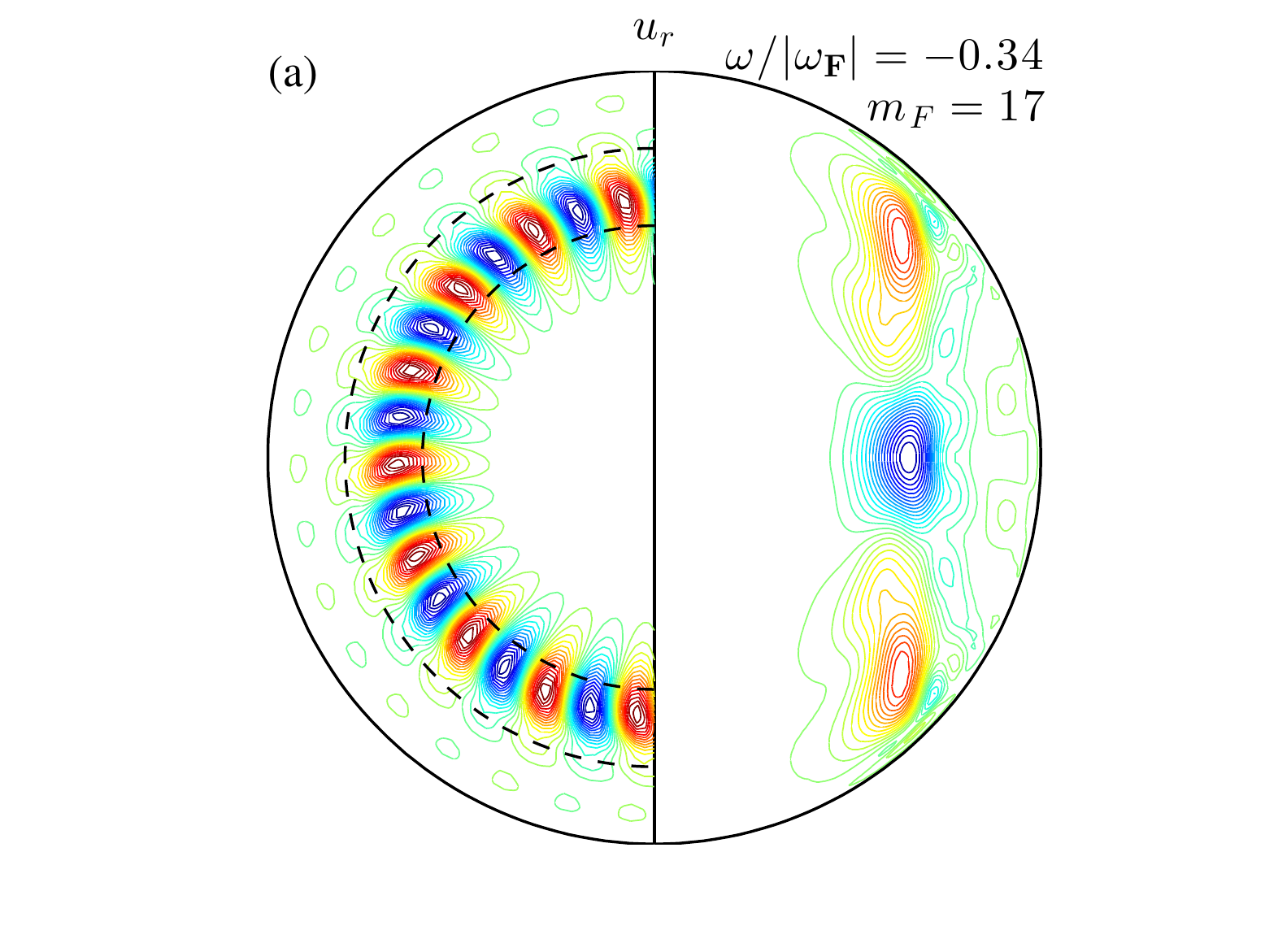}
\includegraphics[width=7cm,clip,trim=2cm 1cm 2cm 0cm]{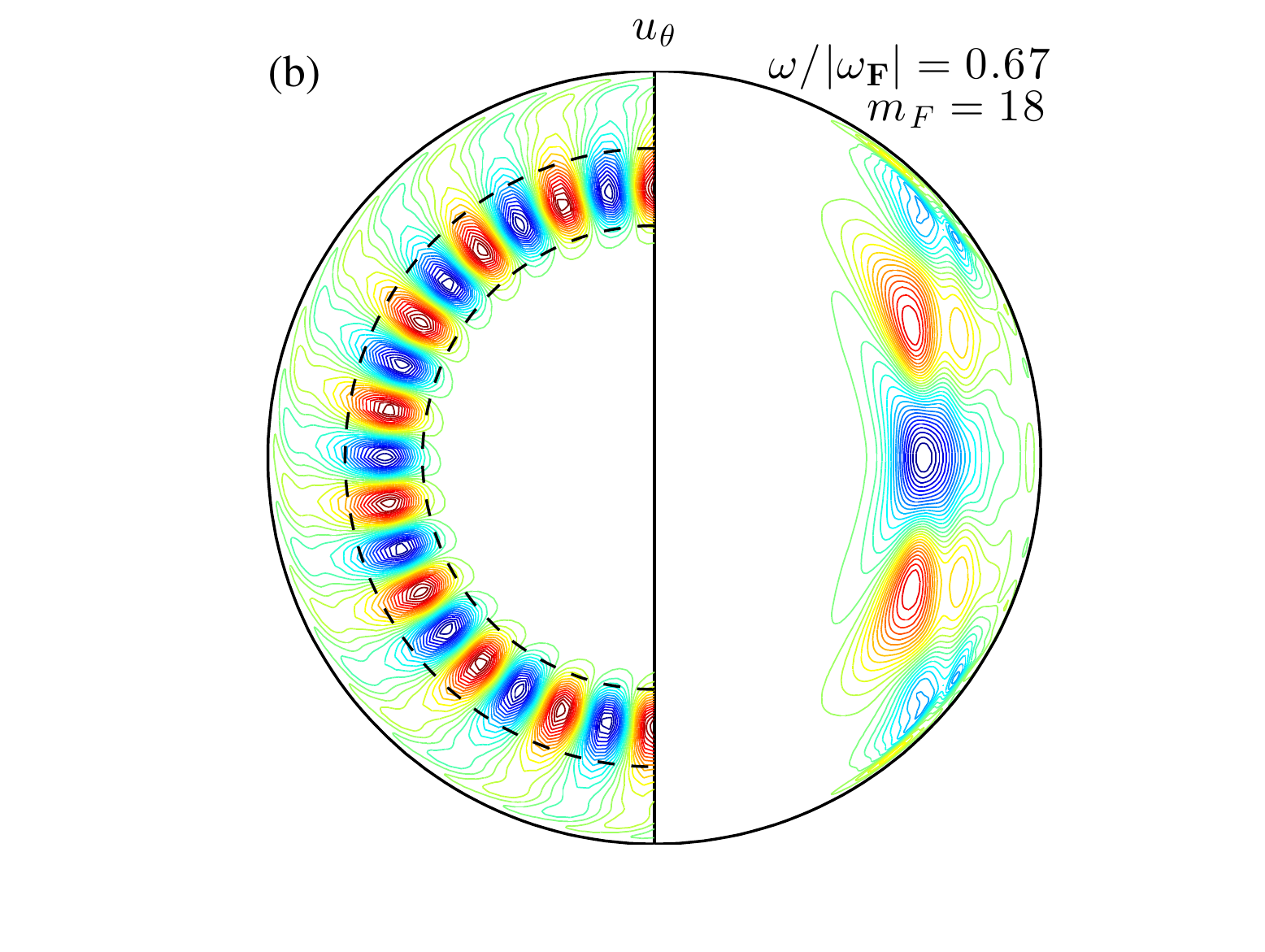}\\
\includegraphics[width=7cm,clip,trim=2cm 1cm 2cm 0cm]{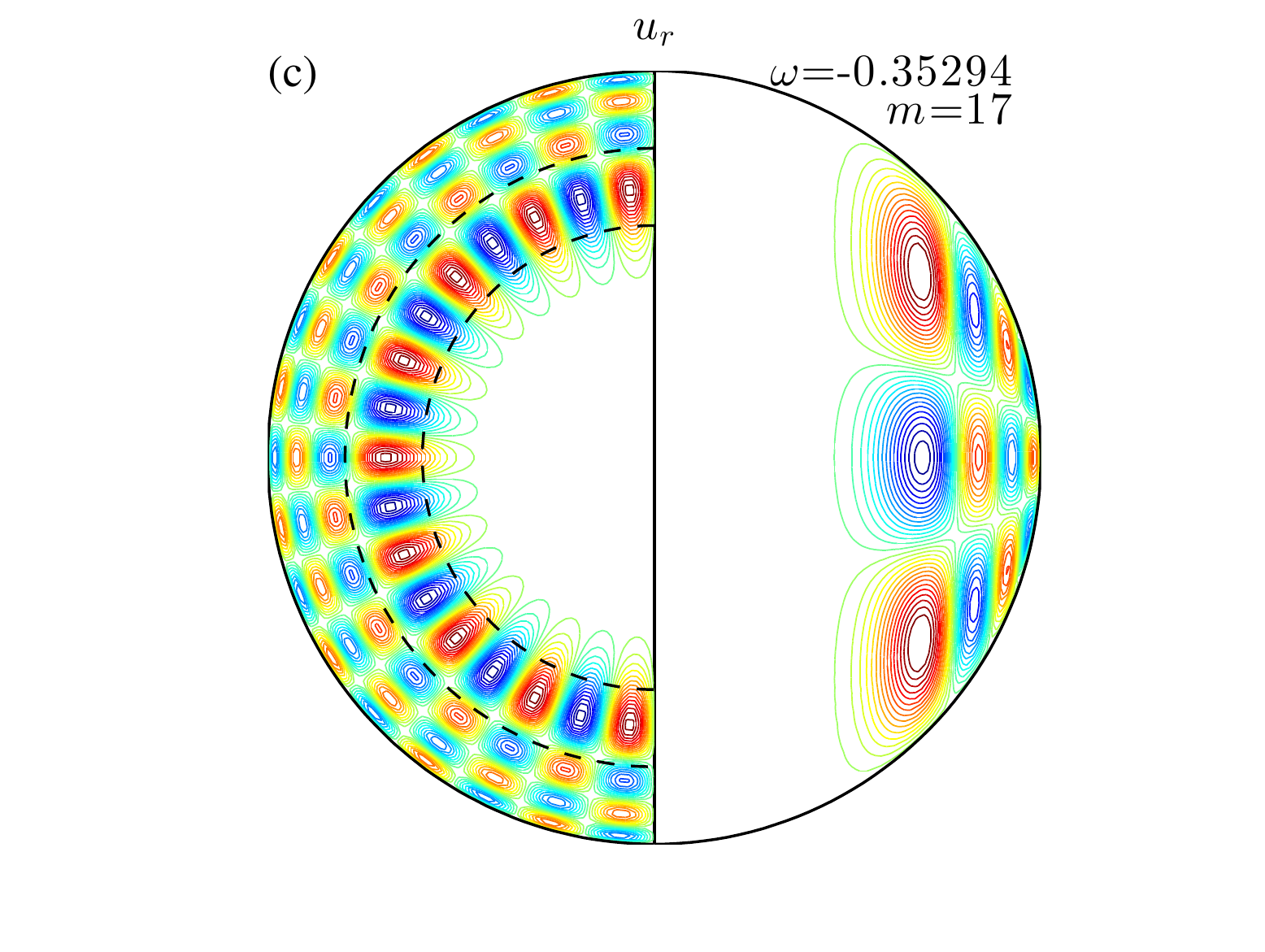}
\includegraphics[width=7cm,clip,trim=2cm 1cm 2cm 0cm]{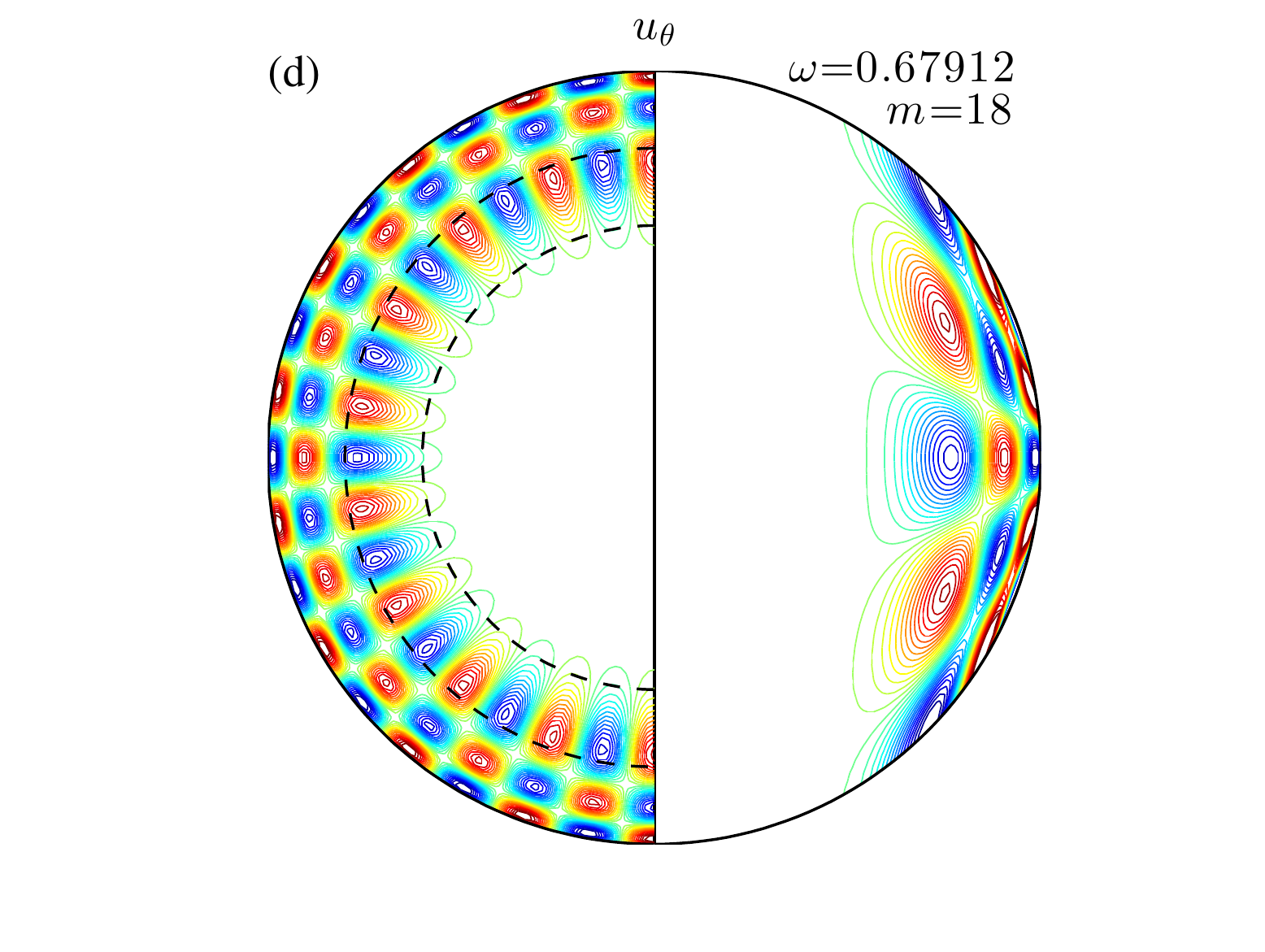}
\end{center}
\caption{Contours of velocities of the unstable modes in the numerical simulations (a-b) and possible inviscid inertial modes (c-d). Snapshots of $u_{r}$ of the $m_F=17$ mode (a) and $u_{\theta }$ of the $m_F=18$ mode (b) are taken at the same instant as in figure \ref{fig:ua_E3e-5P7e-3}. In each plot, the left half is in the equatorial plane and the right half is in a meridional plane. The red (blue) contours represent positive (negative) values. Two dashed lines in the equatorial plane represent two circles with radius of 0.6 and 0.8.}
\label{fig:inviscid_mode}
\end{figure}

The observed azimuthal wavenumbers and frequencies are consistent with a parametric resonance mechanism similar to the shear instability described by Kerswell \cite{Kerswell1993} in precessing spheroids. Such an instability can arise when the background flow couples with two free inertial modes that satisfy the so-called parametric resonant conditions, $\omega_2-\omega_1=1.0$, $m_2-m_1=1$ and $l_2=l_1$, where $\omega$ is the eigen-frequency, $m$ is the azimuthal wavenumber and  $l$ is the spherical harmonic degree of the inertial modes (see Appendix \ref{App1}). To fully characterize the possible inertial modes interacting in our simulations, we calculate the eigen-frequencies and the velocity structures of the inviscid inertial modes in a sphere following the analytical approach in Ref. \onlinecite{Greenspan1968} (see also  Appendix \ref{App1}).
Among all possible inertial modes we only have to consider the antisymmetric ones (odd $l$) with azimuthal wavenumber $m=17$ and $m=18$ based on our observations. In figure \ref{fig:eig_freq} we plot the eigen-frequencies $\omega_{17}+1.0$ and $\omega_{18}$ as a function of $l$ for inertial modes with $m=17$ ($\times$) and and $m=18$ ($\circ$). 
We note that there are several combinations nearly satisfying $\omega_{18}-\omega_{17}=1.0$, i.e. collocated symbols in figure \ref{fig:eig_freq}.
However, only one combination (in the dashed ellipse), with $l=27$, is found to match the observed frequency (the dashed line).         

Figure \ref{fig:inviscid_mode} compares the velocity structures of the observed unstable modes (a-b) in the numerical simulations with the identified inviscid inertial modes (c-d) matching the observations.
% Symmetry considerations allow us to only compare the radial velocities for the $m=17$ mode and the $\theta$-component for the $m=18$ in a meridional plane and in the equatorial plane. 
We observe that the inviscid inertial modes have the same structure in the unstable region $0.6<s<0.8$  as in our numerical simulations. In the most outer region the agreement is more qualitative as we do not observe significant flows in our simulations while the inviscid modes exhibit a well defined pattern with significant velocities. This discrepancy may be attributed to viscous effect that are more influential is this region. 

\subsection{Transition to turbulence}
\begin{figure}
\begin{center}
\includegraphics[width=0.9 \textwidth]{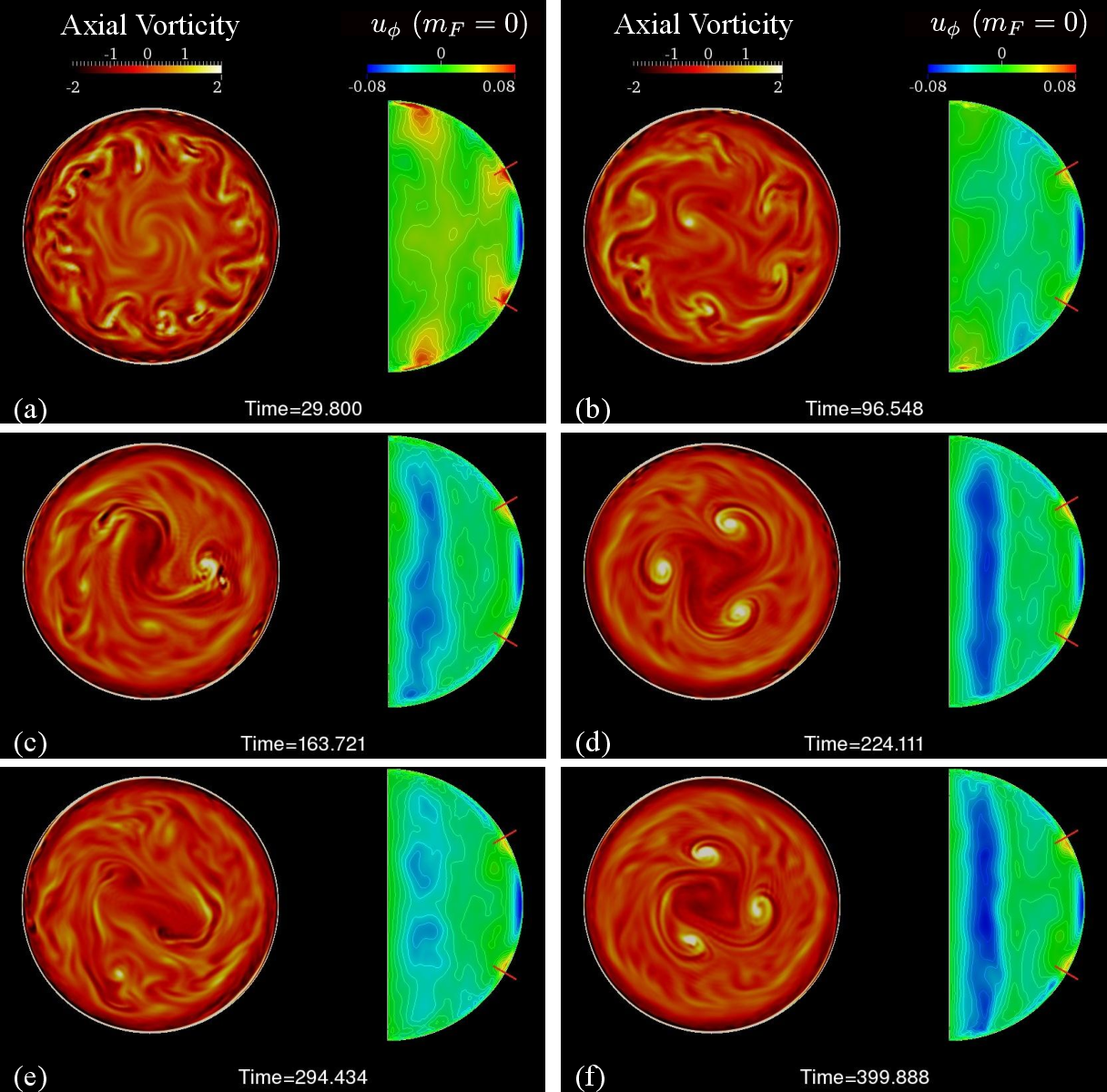}
\includegraphics[width=0.7 \textwidth]{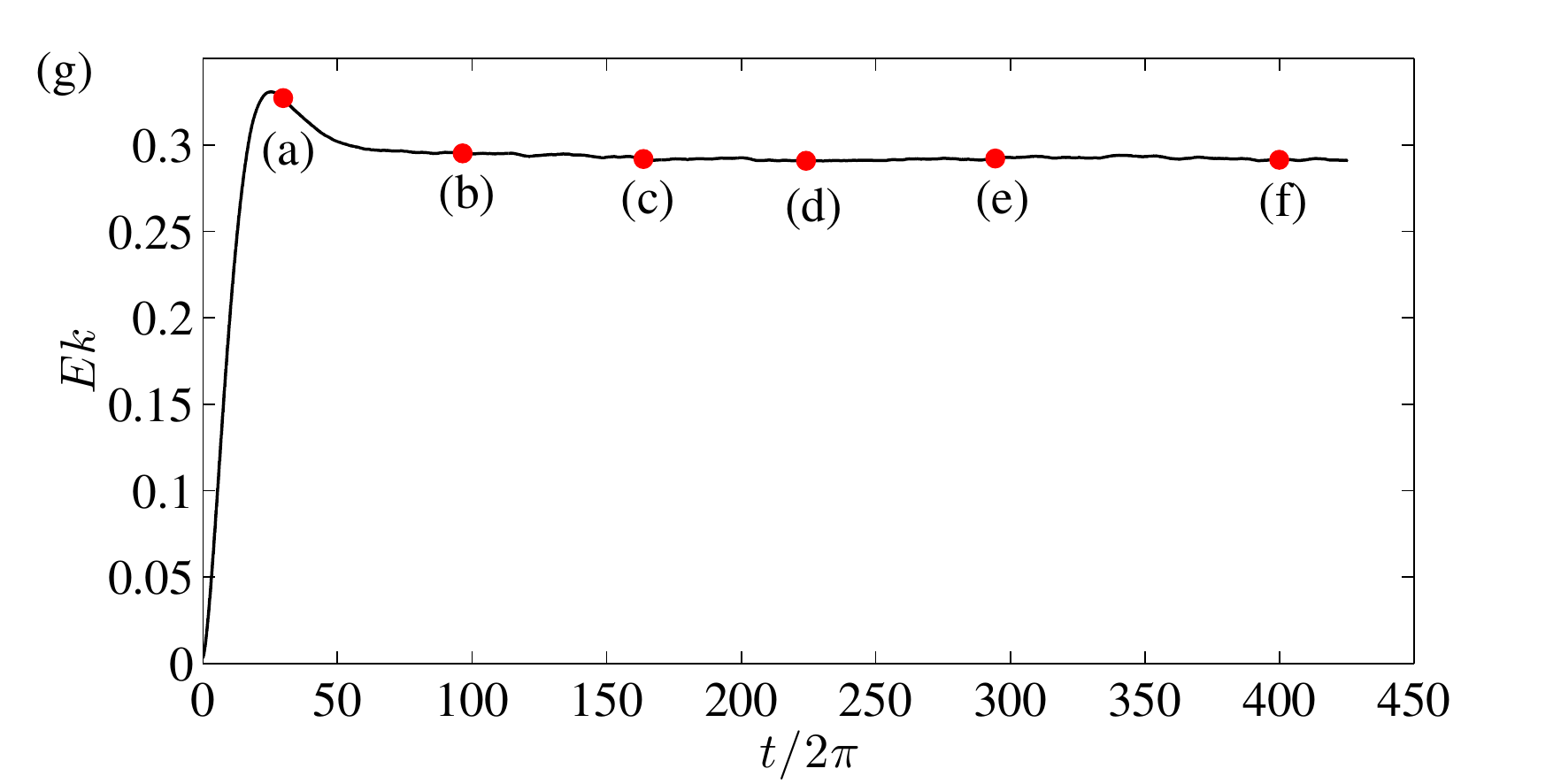}
\caption{(a-f) Snapshots of the axial vorticity in the equatorial plane and axisymmetric azimuthal velocity in a meridional section in the fluid frame. (g) Total kinetic energy $Ek$ in the mantle frame as a function of time. Red dots correspond to snapshots in (a-f). $P_o=-1.35\times 10^{-2}$, $E=3.0\times 10^{-5}$.  }
\label{fig:transition}
\end{center}
\end{figure}
As we increase the precession rate well above threshold, the flow exhibits an inverse cascade of the vorticity. This is illustrated in Figure \ref{fig:transition}, which shows a sequence of snapshots of the axial vorticity in the equatorial plane and the axisymmetric azimuthal velocity in a meridional section in the fluid frame at $P_o=-1.35\times 10^{-2}$ and $E=3.0\times 10^{-5}$. At the beginning (figure \ref{fig:transition} (a)) we see small scale vortices similar to the flow observed during the growth of the parametric instability in figure \ref{fig:ua_E3e-5P7e-3}. The small scale structures then start to merge into large scale elongated cyclonic structures closer to the rotation axis of the fluid (figure \ref{fig:transition} (b-d)). In addition we observed a retrograde (westward) drift of the pattern, as seen in the animation (Movie 2 in supplemental material \cite{Note1}).  
These observations are consistent with an inverse cascade of energy characteristic of two dimensional turbulence. Although not yet formally established, the retrograde drift seems to result from the conservation of angular momentum. %The possible link between these results and the observed westward drift of the earth's magnetic field calls for further investigations.  

In the case of figure \ref{fig:transition}, the large scale cyclones break down to small scales and form again (figure \ref{fig:transition} (e,f)), in some cases the cyclones can be sustained for more than 200 rotation periods of the container. Similar large scale vortices were also observed experimentally in a precessing cylinder \cite{Mouhali2012} and numerically in rotating Rayleigh B\'enard convection \cite{Rubio2014}, yet the underlying merging mechanism remains poorly understood.

\subsection{Stability diagram}
\begin{figure}
\begin{center}
\includegraphics[width=10cm]{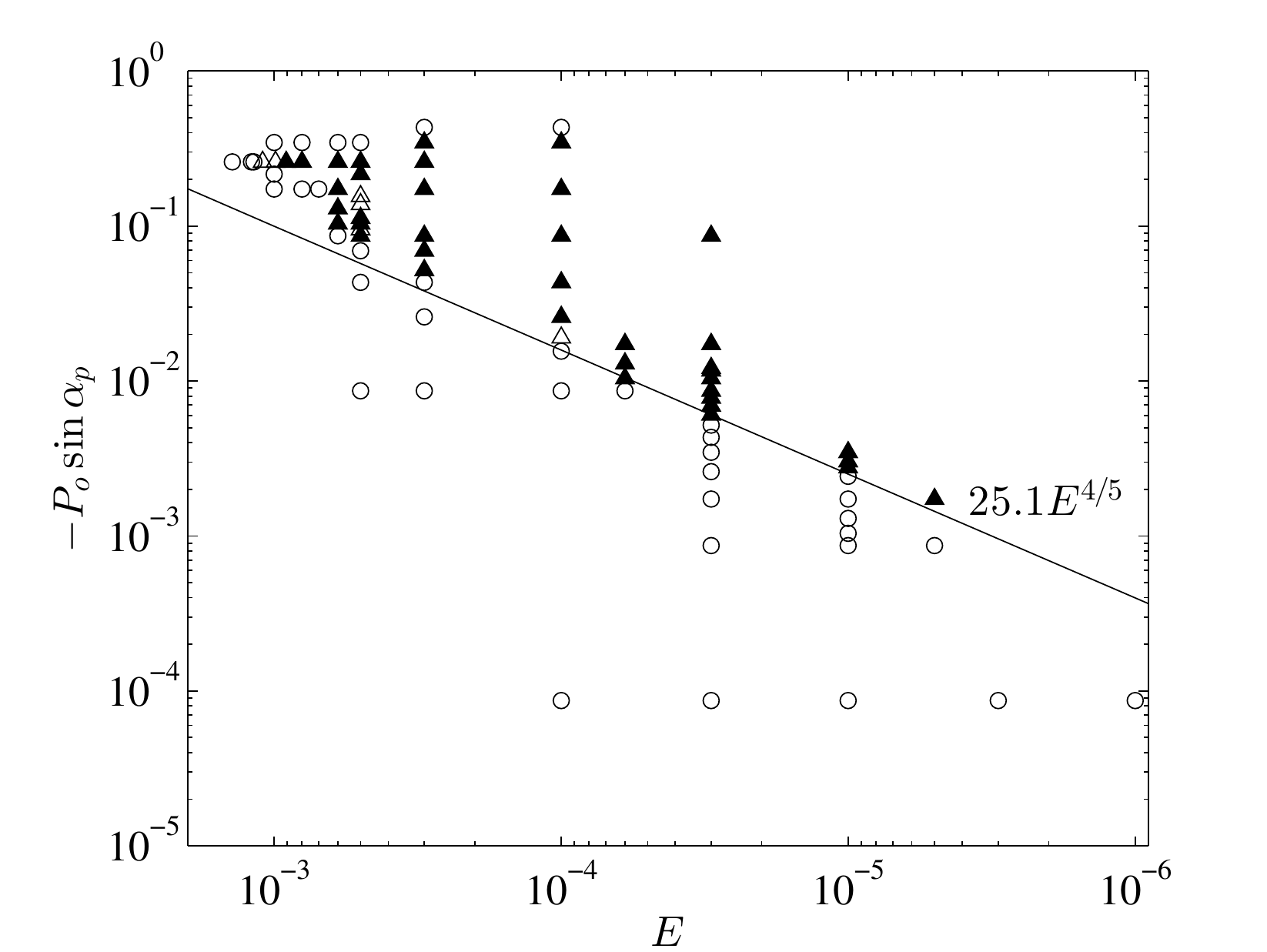}
\caption{Stability diagram in the plane of ($P_o, E$). Circle and triangle symbols represent cases with steady and unsteady kinetic energy, respectively. Open symbols represent flows with centro-symmetry and filled symbols represent symmetry broken flows. The solid line represents the instability threshold from laboratory experiments in a precessing sphere \citep{Kida2013}. }
\label{fig:diagram}
\end{center}
\end{figure}

Figure \ref{fig:diagram} shows the regime diagram in the $(P_o, E)$-parameter space accessible in our numerical study. The flow is characterized as stable (open circles) when it is centrosymmetric and has a steady total kinetic energy after the transient stage. In contrast, symmetry-broken flows (filled triangles) or centrosymmetric flows but with time-varying total kinetic energy (open triangles) indicates the presence of an instability. We note that, although the unstable flows are antisymmetric for the most part, we observe a few cases where the velocity field remains centrosymmetric, as for instance at $P_o=-0.06$ and $E=10^{-4}$. We also observe that the flow in a precessing sphere can be stable when the precession is sufficiently strong, as previously reported in laboratory experiments \citep{Goto2007, Goto2014}. In these cases, the fluid axis is nearly aligned with the precession vector.

The solid line in figure \ref{fig:diagram} represents the lower instability threshold of laboratory experiments in a precessing sphere \citep{Kida2013}. We can see that our numerical simulations are in good agreement with the experimental results particularly when $E\leq 10^{-4}$. Finally, both numerical and experimental results are consistent with a critical Poincar\'e number scaling as $|P_o|\sin \alpha_p=O(E^{4/5})$ that we shall now derive on the basis of a parametric instability mechanism.       

\section{The conical shear-driven parametric instability (CSI)} \label{Sec:Scal}

Our observations are consistent with a parametric instability mechanism similar to the shear instability described by Kerswell \cite{Kerswell1993}. However, in contrast with the spheroidal geometry the shear cannot result from  topographic effects. Instead, we argue that the conical shear layers spawned from the critical latitudes in the boundary layers can also induce a parametric instability. Indeed, the conical shear layers driven by the linear viscous interactions in the boundary layer are $m_F=1$ and $\omega/|\bm{\omega_F|}=1.0$, in the fluid frame, and steady in the precession frame. Thus, they satisfy the parametric resonant conditions with the observed modes in our simulations. Hereinafter we will refer to this instability as a CSI, for Conical-Shear-Instability.

\begin{figure}
\begin{center}
\includegraphics[width=0.9\textwidth]{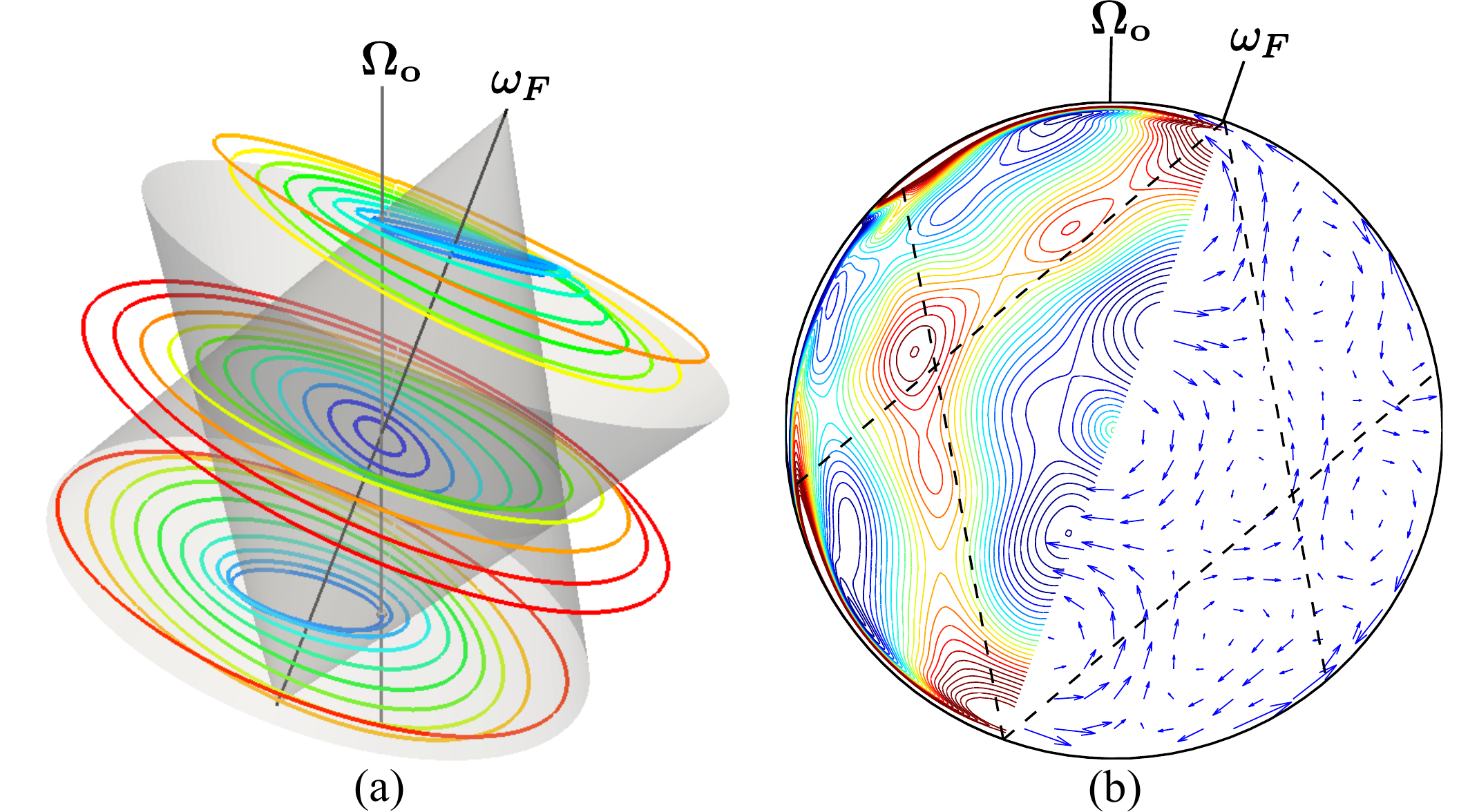}
\end{center}
\caption{(a) Streamlines of the velocities in the precession frame at $t/2\pi=200$. The gray line represents the rotation axis of the mantle and the black line represents the rotation axis of the fluid. Shadowed conical cones represent the characteristic surfaces with respect to the rotation axis of the fluid. (b) Contours of $u_{\theta}$ (left half) and in-plane velocity vectors (right half) in the meridional plane across both the rotation axis of fluid and the rotation axis of the fluid at the same instant as in (a). Dashed lines are intersections of the characteristic surfaces and the meridional plane. $P_o=-7.0\times 10^{-3}$, $E=3.0\times 10^{-5}$. }
\label{fig:streamline}
\end{figure}

Figure \ref{fig:streamline} (a) shows the streamlines in the frame of precession just before the growth of the unstable modes illustrating the distortion along the conical shear surfaces, here represented in light grey. Figure \ref{fig:streamline} (b) shows the traces of the conical shear layers (left half) and velocity vectors (right half) in the meridional plane ($\bm{\Omega_o},\bm{\omega}$). 

Following the well established theory \cite{Kerswell2002, Lacaze2004}, we shall now derive the scaling for the onset condition. For an exact resonance, the growth rate of the parametric instability can be written as
\begin{equation}
\label{eq:sigma}
\sigma=\sqrt{C_1C_2}-\frac{(\kappa_1+\kappa_2)}{2}E^{1/2},
\end{equation} 
where $\kappa_1$ and $\kappa_2$ are the real decay rate factors of two inertial modes. The interaction parameters between the conical shear $\bm u_{shear}$ and two inertial modes  $\bm{u}_1$, $\bm u_2$  are given by
\begin{equation}
\label{eq:C1}
C_1=\frac{<\bm u_1, \bm u_2\times(\nabla\times \bm u_{shear})>+<\bm u_1,\bm u_{shear}\times(\nabla\times \bm u_2)>}{<\bm u_1, \bm u_1>},
\end{equation}  
\begin{equation}\label{eq:C2}
C_2=\frac{<\bm u_2, \bm u_1\times(\nabla\times \bm u_{shear})>+<\bm u_2,\bm u_{shear}\times(\nabla\times \bm u_1)>}{<\bm u_2, \bm u_2>},
\end{equation}  
with $<\bm A, \bm B>=\int\int\int_V \bm A^*\cdot 
\bm B\,\mathrm d V$ and $\bm A^*$ is the complex conjugate of $\bm A$.

A complete derivation of the growth rate requires evaluation of the volume integrals in Equations. (\ref{eq:C1},\ref{eq:C2}). Although an explicit expression of $\bm u_{shear}$ was derived recently \citep{Kida2011}, a set of partial differential equations have to be solved numerically to get $\bm u_{shear}$. Nevertheless, a scaling law for the onset of the parametric instability can be established using heuristic arguments. It has been demonstrated that the amplitude of the conical shear layers scales as $|\bm u_{shear}|=O(\varepsilon E^{1/5})$ over a width $O(E^{1/5})$, where $\varepsilon$ represents the differential rotation between the fluid and the surrounding solid shell \cite{Stewartson1963,Noir2001,Kida2011}. Therefore, we estimate $|\nabla\times \bm u_{shear}|=O(\varepsilon E^{1/5}/E^{1/5})=O(\varepsilon)$, while the integration volume is proportional to $E^{1/5}$. It follows,
\begin{equation}
\label{eq:C1C2}
\sqrt{C_1C_2}=O(\varepsilon E^{1/5}).
\end{equation}  
Combining Eq.(\ref{eq:C1C2}) and Eq. (\ref{eq:sigma}), we obtain the instability threshold (Re($\sigma)=0$):
\begin{equation}
\label{eq:threshold0}
\varepsilon E^{1/5} \sim \frac{(\kappa_1+\kappa_2)}{2}E^{1/2},
\end{equation}
leading to
\begin{equation}
\label{eq:threshold1}
\varepsilon = O(E^{3/10}).
\end{equation}
In a precessing sphere, the differential rotation $\varepsilon$ is a function of $P_o$, $\alpha_p$ and $E$. At low precession rate, i.e. $|P_o|\sin \alpha_p\ll E^{1/2}$, the direct resonance mechanism between the tilt-over mode and the precessional forcing leads to $\varepsilon=O(|P_o|\sin \alpha_p/E^{1/2})$ \cite{Busse1968, Kida2011}. This scaling is also confirmed by our numerical simulations in figure \ref{fig:alpha_f}. Thus the lower bound for the threshold of the CSI in a sphere can be written
%In this range of low Poincar\'e number the threshold for the SCI in a sphere can be written 
\begin{equation}
|P_o|\sin \alpha_p=O(E^{4/5}).
\end{equation}
This scaling is in quantitative agreement with our numerical simulations (figure \ref{fig:diagram}) and previous experimental results \cite{Kida2013}.

The conical shear layers can also be excited in precessing spheroidal cavities \cite{Tilgner1999a} and thus the CSI should be induced as well. In this geometry, we must calculate the differential rotation $\varepsilon$ following Busse \cite{Busse1968} which also depends on the ellipticity $\eta$ of a spheroid,  and then apply the instability criterion of equation \ref{eq:threshold1}.

\section{Discussion and concluding remarks} \label{Sec:Diss}
\begin{figure}
\begin{center}
\includegraphics[width=16cm,]{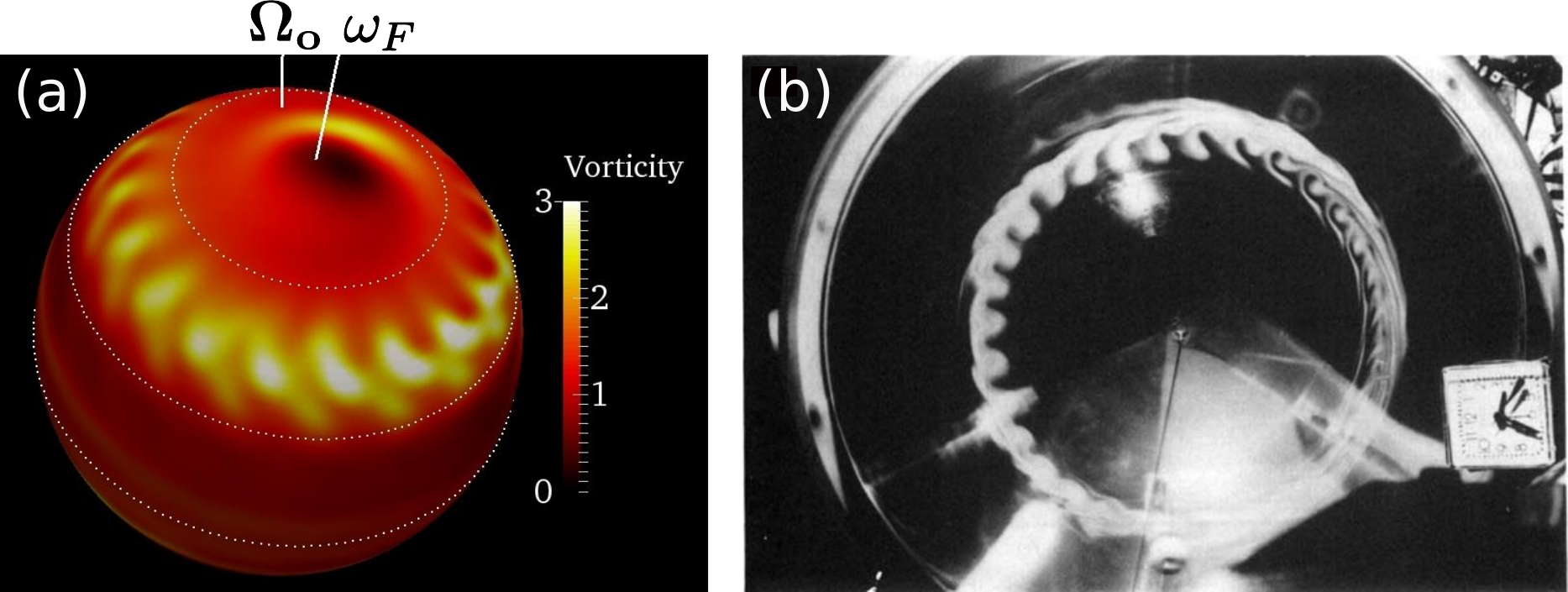}
\end{center}
\caption{(a) Snapshot of total vorticity on the surface at $r=1-10 E^{1/2}$ in the precession frame at the same instant as figure \ref{fig:ua_E3e-5P7e-3} . Three white dotted lines represent the latitudes of $0^\circ$, $30^\circ$ and $60^\circ$ with respect to the rotation axis of the fluid.  $P_o=-7.0\times10^{-3}$, $E=3.0\times 10^{-5}$. (b) Advection of dye injected in the boundary layer of a precessing spheroid (Reprinted with permission from J. Vanyo, P. Wilde, P.Cardin, and P. Olson, Geophys. J. Int. 121, 136(1995). Copyright 1995, Oxford University Press). $\eta=1/100$, $\alpha_p=23.5^{\circ}$, $P_o=-5.0\times 10^{-4}$ and $E=8.0\times10^{-7}$.}
\label{fig:Vanyo1995}
\end{figure}

\begin{table}
\caption{Comparison of different instability mechanisms in spheroids used in experiments by Vanyo et. al. \cite{Vanyo1995} and Malkus \cite{Malkus1968}  and in planetary liquid cores. The growth rates of the first two mechanisms are given by Kerswell \cite{Kerswell1993} and the CSI discussed in this study. The prefactors of the scaling laws are assumed to be around $O(1)$. The differential rotation $\delta\omega$ is calculated using Busse's  theory \cite{Busse1968} for the spheroid. }
\label{tab:compare}
\begin{ruledtabular}
\begin{tabular}{llcccc}
 & Parameter 		&  Malkus		& Vanyo		& Earth 						& Moon  \\
Ellipticity &  $\eta $		&  $4.0\times 10^{-2}$			&  $1.0\times 10^{-2}$ 				& $\sim2.5\times 10^{-3}$ 					& $\sim10^{-5}$  \\

Ekman number & $E$ 		& $1.0\times10^{-5}$					& $8.0\times10^{-7}$ 		& $\sim10^{-14}$ 			& $\sim10^{-12}$ \\

Differential rotation &  $\varepsilon$ 	& $3.3\times 10^{-1}$ 		& $2.0\times 10^{-2}$ 					& $\sim 1.7\times 10^{-5}$ 	& $\sim 3.0\times 10^{-2}$  \\

\hspace*{0.2cm} \\
 & Inviscid growth rate  &   &  &\\
Elliptical Instability  & $O(\varepsilon^2 \eta) $ & $\sim4.4\times 10^{-3}$  & $\sim4.0\times 10^{-6}$   &  $\sim7.2\times 10^{-13}$ & $\sim 9.0\times 10^{-9}$\\
Shear Instability & $O(\varepsilon \eta) $ & $\sim1.3\times 10^{-2}$ & $\sim2.0\times 10^{-4}$ &  $\sim4.3\times 10^{-8}$ & $\sim3\times10^{-7}$ \\
CSI & $O(\varepsilon E^{1/5})$& $\sim3.3\times 10^{-2}$ & $\sim1.2\times 10^{-3}$  & $\sim2.7\times 10^{-8}$ & $\sim1.2\times10^{-4}$  \\
\hspace*{0.2cm}\\  
 & Viscous decay rate  &   &  &\\
Viscous damping & $O(E^{1/2})$ & $\sim 3.2\times 10^{-3}$ & $\sim 8.9\times 10^{-4}$ & $\sim 10^{-7}$  & $\sim 10^{-6}$\\
\end{tabular}
\end{ruledtabular}
\end{table}

The present numerical study investigates the stability of  precession-driven flows in a full sphere at moderate to low Ekman numbers. It is found that at low precession rate the flow is of uniform vorticity with a viscous correction superimposed. As the precession rate increases, the internal conical shear layers driven by the linear interactions in the boundary layer induce a parametric instability. The threshold conditions have been established using heuristic arguments leading to a scaling $|P_o|\sin \alpha_p=O(E^{4/5})$ in quantitative agreement with both our numerical simulations and former experimental observations. At onset, the inertial modes involved in the destabilization mechanism concentrate the energy in an annular region between two cylinders coaxial with the rotation axis of the fluid ($0.6<s<0.8$). Above threshold, the flow evolves from its onset geometry with relatively large azimuthal wave numbers to a simpler structure with few cyclonic vortices, closer to axis of rotation of the fluid and traveling in a westward direction.  

Our results may allow us to shed light on the long standing ill-understood wave-like instabilities reported by Vanyo et. al. \cite{Vanyo1995} almost 20 years ago in a precessing spheroid  with ellipticity $\eta=1/100$. In one of their experiments dye was injected next to the boundary layer of a precessing fluid cavity. Pictures taken over more than one hour of experimentation reveal travelling wave-like structures organized on a circular path concentric with the rotation axis of the fluid. Figure \ref{fig:Vanyo1995} compares the observations from dye injection by Vanyo et. al. \cite{Vanyo1995} with the vorticity in our numerical simulations. We argue that the spiraling structure observed by Vanyo et.al \cite{Vanyo1995} can be explained as a pair of traveling inertial modes excited through a parametric resonance. In Vanyo's experiments, which used a spheroid, three mechanisms may generate a parametric instability: the two mechanisms proposed by Kerswell \cite{Kerswell1993} and the conical shear layer driven one investigated in the present paper. To establish the dominant mechanism, we first calculate the differential rotation $\varepsilon$ using Busse's theory \cite{Busse1968} in the spheroid of Vanyo et. al.\cite{Vanyo1995}, then we use Eq.\ref{eq:threshold1} for the CSI and the derivation of Kerswell \cite{Kerswell1993} for the two other mechanisms. The results summarized in Table \ref{tab:compare} clearly show that the experimental setup of Vanyo et. al. \cite{Vanyo1995} is more prone to CSI rather than the classical topography driven parametric instabilities of Kerswell \cite{Kerswell1993}. In the original experiment in precessing spheroid by Malkus \cite{Malkus1968}, the wave like instability was also reported (see his Figure 2 (b)). Based on our estimates in Table \ref{tab:compare}, both the topographic effect and the conical shear layers may lead to the parametric instability in the experiment by  Malkus \cite{Malkus1968}.

Finally, we shall discuss our results in a planetary context. Using a hydrostatic model for the shape of the Lunar's Core-Mantle boundary (CMB) the differential rotation $\varepsilon$ due to the 18.6 years precession has been estimated to be of order $3\%$ of the mantle rotation \cite{Williams2001,Noir2013}. In line with this conclusion, Williams et. al.\cite{Williams2001} argued that the resulting flow should be turbulent in order to account for the large dissipation inferred from the Lunar Laser Ranging (LLR) measurements. Yet, if one considers the two classical precessional instabilities of Kerswell \cite{Kerswell1993} due to the polar ellipticity, the inviscid growth rate remains smaller than the viscous decay rate leading to a stable flow. In contrast, considering the effects of the conical shear layers we obtain an inviscid growth rate 100 times larger than the viscous decay rate assuming that the Ekman number is around $10^{-12}$(Table \ref{tab:compare}). These highly super critical conditions are likely to be associated with more complex flows than the one reported here at moderate Ekman number simulations, leading to significant dissipation as observed through inversion of the LLR time series.

In contrast with the Moon, the Earth's polar flattening ($\eta\approx 1/400$ on the CMB \cite{Vanyo1995}) and the large period of precession (26 000 years) result in a liquid core strongly coupled to the mantle. As a consequence, the differential rotation is small at present day, $\varepsilon\approx 1.7\times 10^{-5}$. As seen in Table \ref{tab:compare}, all three mechanisms may lead to a stable precession-driven flow in the outer core assuming the Ekman number of the outer core is around $10^{-14}$. %However, the presence of the inner core may affect the conclusion drawn here. 
In addition, it has been shown that another set of conical shear layers emanate from the inner core boundary \cite{Hollerbach1995,Kerswell1996}, with an amplitude of $O(\varepsilon E^{1/6})$ over a thickness of $O(E^{1/3})$. In principle these conical shear layers could also participate to a CSI. A stability criterion can then be derived using the same heuristic argument presented herein leading to $\varepsilon= O(E^{1/3})$. Hence, CSI in the earth's liquid core may occur if the Ekman number of the outer core is below $10^{-15}$.  
%Unfortunately, the differential rotation between the inner core and the outer core is difficult to estimate  

%In contrast with the Moon, the Earth's polar flattening ($\eta\approx 1/400$ on the CMB \cite{Vanyo1995}) and the large period of precession (26 000 years) result in a liquid core strongly coupled to the mantle. As a consequence, the differential rotation is small at present day, $\varepsilon\approx 1.7\times 10^{-5}$. As seen on table \ref{tab:compare}, all three mechanisms may lead to a stable flow in outer core assuming the Ekman number of the outer core is around $10^{-14}$. However, the presence of the inner core may affect the conclusion drawn here. Indeed, it has been shown that another set of conical shear layers emanate from the inner core boundary \cite{Hollerbach1995,Kerswell1996}, with an amplitude of $O(\varepsilon E^{1/6})$ over a thickness of $O(E^{1/3})$. In principle these conical shear layers could also participate to a CSI. A stability criterion can then be derived using the same heuristic argument presented herein leading to $\varepsilon= O(E^{1/3})$. 
%  for which it would require that the Ekman number of the Earth's outer core is smaller than $10^{-15}$ to be influential. 

In planetary condition, however, several others parameters have to be considered to draw a more reliable conclusion such as the effect of stratification and magnetic fields, as well as the interaction of precession driven flows with other sources of motions such as thermo-chemical convection.

\begin{acknowledgments}
We would like to acknowledge Andrew Jackson for very useful discussions and suggestions on this study. We thank Andreas Fichtner for computational assistance. Simulations were run on Swiss National Supercomputing Center (CSCS) under the project s369. Figure 13 (b) is reproduced from the Figure. 5(b) in Ref. \onlinecite{Vanyo1995}, for which we thank the Oxford University Press and the original authors for permission to reprint it. YL and JN are supported by ERC Grant No. 247303 (MFECE) at ETH Zurich. PM is supported by the National Science Foundation EAR CSEDI \#1067944 grant. 
\end{acknowledgments}

\appendix 
\section{Symmetry of the base flow}\label{app_symmetry}
The linear viscous solution of the governing Eqs. (\ref{eq:NS1}-\ref{eq:NS2}) is symmetric around the origin, namely $\bm u(\bm r)=-\bm u(-\bm r)$. Neglecting the nonlinear term in Eq. (\ref{eq:NS1}) and changing the sign of the position vector, we have 
\begin{equation}
\frac{\partial \bm u(\bm r)}{\partial t}+2(\bm{\hat{k}}+P_o \bm{\hat{k}_p})\times \bm u(\bm r)=-\nabla p(\bm r)+E\nabla^2\bm u(\bm r)-P_o(\bm{\hat{k}_p}\times \bm{\hat{k}})\times \bm r,
\label{eq:app1}
\end{equation}
and 
\begin{equation}
\frac{\partial \bm u(-\bm r)}{\partial t}+2(\bm{\hat{k}}+P_o \bm{\hat{k}_p})\times \bm u(-\bm r)=\nabla p(-\bm r)+E\nabla^2\bm u(-\bm r)+P_o(\bm{\hat{k}_p}\times \bm{\hat{k}})\times \bm r.
\label{eq:app2}
\end{equation}
Adding and subtracting Eqs. (\ref{eq:app1}) and (\ref{eq:app2}), we obtain 
\begin{equation}
\frac{\partial \bm u_a}{\partial t}+2(\bm{\hat{k}}+P_o \bm{\hat{k}_p})\times \bm u_a=-\nabla p_a+E\nabla^2\bm u_a,
\label{eq:app3}
\end{equation}
and 
\begin{equation}
\frac{\partial \bm u_s}{\partial t}+2(\bm{\hat{k}}+P_o \bm{\hat{k}_p})\times \bm u_s=-\nabla p_s+E\nabla^2\bm u_s-P_o(\bm{\hat{k}_p}\times \bm{\hat{k}})\times \bm r,
\label{eq:app4}
\end{equation}
where 
\begin{equation}
\bm u_s=\frac{\bm{u}(\bm r)-\bm u(-\bm r)}{2}, \quad p_s=\frac{p(\bm r)+p(-\bm r)}{2},
\end{equation}
and 
\begin{equation}
\bm u_a=\frac{\bm{u}(\bm r)+\bm u(-\bm r)}{2}, \quad  p_a=\frac{p(\bm r)-p(-\bm r)}{2},
\end{equation}
are symmetric and antisymmetric parts respectively. We can see that only the symmetric solution is forced by precession. 
\section{Base flow in the mantle frame and precession frame} \label{App0}
\begin{figure}
\includegraphics[width=0.99\textwidth]{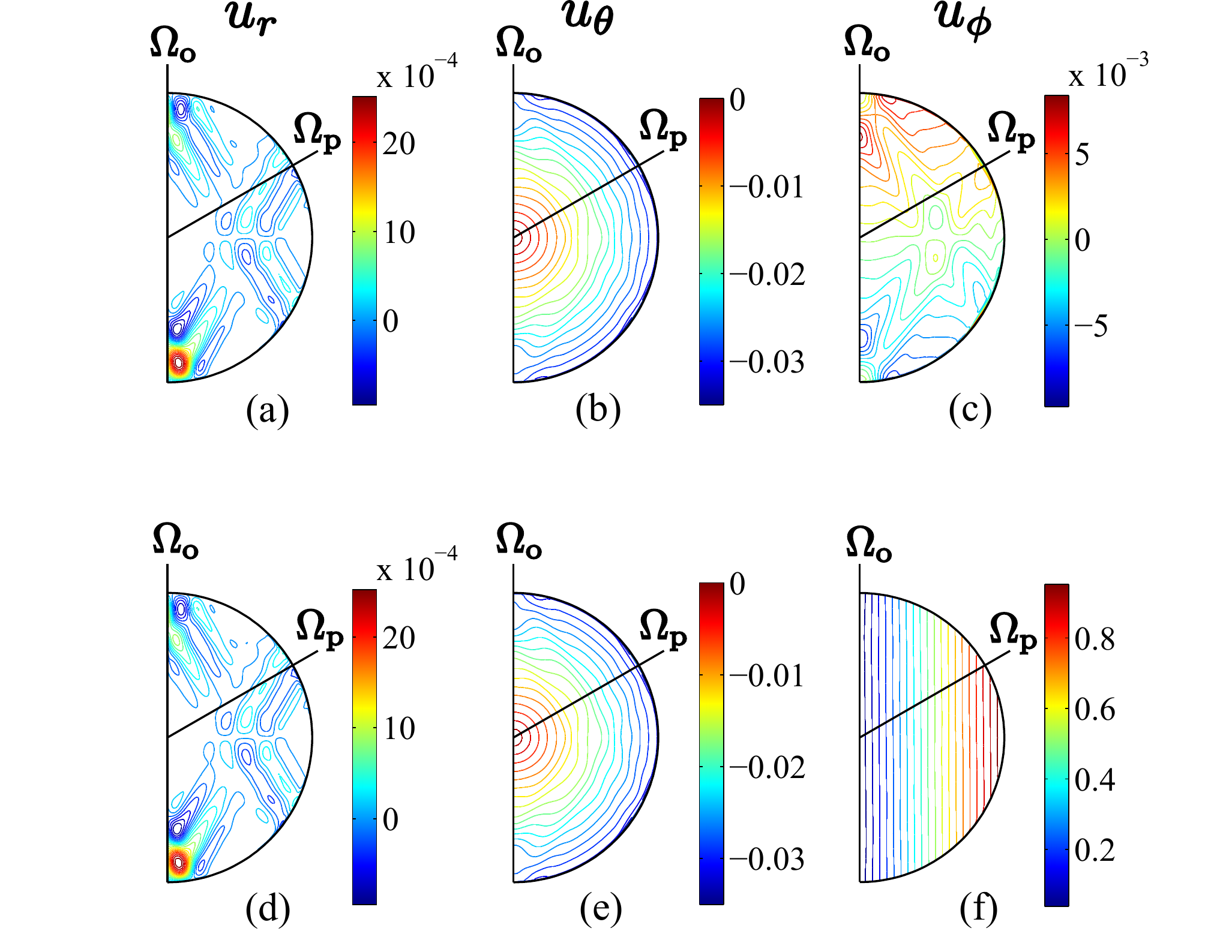}
\caption{Contours of the velocities in the meridional plane across both $\bm {\Omega_o}$ and $\bm {\Omega_p}$ in the mantle frame (a-c) and the precession frame (d-f) at $P_o=-1.0\times 10^{-4}$ and $E=1.0\times 10^{-6}$.}
\label{fig:vel_merid}
\end{figure}

In order to provide different views of precession driven base flow as suggested by one referee, we show the velocities in the mantle frame and the precession frame in Fig. \ref{fig:vel_merid} for $P_o=-1.0\times 10^{-4}$ and $E=1.0\times 10^{-6}$. In the mantle frame, the base flow is a combination of the Poincar\'e mode, i.e. a solid body rotation about an axis in the equatorial plane, and the secondary flow due to the viscous correction. The secondary flow is dominated by the conical shear layers spawned from the critical latitudes, which is hidden behind the solid body rotation. Note that the Poincar\'e mode is a travelling inertial mode in the mantle frame. In the view of the precession frame, we see both global rotation of the container  and the solid body rotation around an axis in the equatorial plane (the Poincar\'e mode), leading to a tilted rotation axis of the fluid with respect to the rotation axis of the container. The flow is steady in the precession frame. Again, the secondary flow due to the viscous correction is hidden beneath the solid body rotation.   

\section{Inertial modes in a sphere} \label{App1}
Inertial modes are solutions of the so-called Poincar\'e equation which can be written as \cite{Greenspan1968}
\begin{equation} 
\frac{1}{s}\frac{\partial p}{\partial s} +\frac{1}{s^2}\frac{\partial^2 p}{\partial s^2}+\frac{1}{s^2}\frac{\partial^2 p}{\partial \phi^2}-\Big(\frac{4-\omega^2}{\omega^2}\Big)\frac{\partial ^2p}{\partial z^2}=0,
\end{equation} 
in the cylindrical coordinates ($s,\phi,z$) with the no-penetration boundary condition
\begin{equation}
s\frac{\partial p}{\partial s}+\frac{2}{\mathrm{i} \omega}\frac{\partial p}{\partial \phi}+(1-\frac{4}{\omega^2}z\frac{\partial p}{\partial z})=0,
\end{equation}
on the spherical surface $s^2+z^2=1$. Solutions can be found using separation of variables and they are
\begin{equation}
p_{klm}(s,\phi,z)=s^mz^\gamma\mathrm{e}^{\mathrm{i}m\phi}\prod_{j=1}^N(x_j^2(x_j^2-1)+x_j^2(1-\frac{\omega_{klm}^2}{4})s^2+\frac{\omega_{klm}^2}{4}(1-x_j^2)z^2),
\end{equation}
where $\gamma=0$ if $(l-m)$ is even and $\gamma=1$ if $(l-m)$ is odd, and $x_j$ are the $N=\frac{1}{2}(l-m-\gamma)$ zeros of the Legendre polynomial $P_l^m(x)$ of degree $l$ and order $m$. The eigen frequency $\omega_{klm}$ is the $k$th solution of the transcendental equation
\begin{equation}
2(1-\frac{\omega^2}{4})\frac{\mathrm d }{\mathrm d \omega}P_l^m(\frac{\omega}{2})=mP_l^m(\frac{\omega}{2}).
\end{equation} 
For each given integer of $l$ and $m$, there are $l-m$ solutions if $m\neq0$ and $l-1$ solutions if $m=0$. Each eigen frequency $\omega_{klm}$ and the associated eigen function are identified by three indexes ($k,l,m$). Here we only consider $m\geqslant 0$ and the solution can be extended to $m<0$ by the relationship $\omega(k,l,-m)=-\omega(k,l,m)$.

Finally, the velocity field of the each eigen mode is obtained from
\begin{equation}
\boldsymbol{u}_{klm}=\frac{\mathrm{i}}{\omega(4-\omega^2)}[4(\boldsymbol{\hat z}\cdot \nabla p_{klm})\boldsymbol{\hat z}-\omega^2\nabla p_{klm}- 2\mathrm{i}\omega \boldsymbol{\hat z}\times \nabla p_{klm}].
\end{equation} 
The explicit expression of velocities are given by Zhang et. al. \cite{Zhang2001}. 

%\bibliography{precession}

%Merlin.mbs v4.21 2009-07-09.
%

\end{document}